\documentclass[float,epsfig,usenatbib]{mn2e}

\usepackage{graphicx}
\usepackage{psfrag}
\usepackage{amsmath,amssymb}
\usepackage{float}
\usepackage{afterpage}
\usepackage{longtable,lscape}
\usepackage{rotating}

\title[xGASS: the extended GALEX Arecibo SDSS Survey]{xGASS: Total cold gas scaling relations and molecular-to- atomic gas ratios of galaxies in the local Universe}

\author[B. Catinella et al.]
{Barbara Catinella,$^{1}$\thanks{barbara.catinella@uwa.edu.au}
Am\'elie Saintonge,$^{2}$ Steven Janowiecki,$^{1}$ Luca Cortese,$^{1}$ 
\newauthor  Romeel Dav\'e,$^{3}$ Jenna J. Lemonias,$^{4}$ Andrew P. Cooper,$^{5}$ David Schiminovich,$^{4}$
\newauthor  Cameron B. Hummels,$^{6}$ Silvia Fabello,$^{7}$ Katinka Ger\'eb,$^{8}$ Virginia Kilborn$^{8}$
\newauthor  and Jing Wang$^{9}$ \\
$^{1}$International Centre for Radio Astronomy Research, M468, The University of Western Australia, Crawley, WA 6009, Australia\\
$^{2}$Department of Physics \& Astronomy, University College London, Gower Place, London WC1E 6BT, UK\\
$^{3}$Department of Physics \& Astronomy, University of the Western Cape, Bellville, Cape Town 7535, South Africa \\
$^{4}$Department of Astronomy, Columbia University, New York, NY 10027, USA\\
$^{5}$Institute for Computational Cosmology, Department of Physics, University of Durham, South Road, Durham DH1 3LE, UK \\
$^{6}$TAPIR, California Institute of Technology, Pasadena, CA 91125, USA \\
$^{7}$Autoliv Electronics Germany, Theodor-Heuss-Str. 2, 85221 Dachau, Germany\\
$^{8}$Centre for Astrophysics and Supercomputing, Swinburne University of Technology, Hawthorn, VIC 3122, Australia\\
$^{9}$Kavli Institute for Astronomy and Astrophysics, Peking University, Beijing 100871, China\\
}

\date{}

\begin{document}

\def\deg{$^{\circ}$}
\newcommand{\eg}{{\it e.g.}}
\newcommand{\ie}{{\it i.e.}}
\newcommand{\minusone}{$^{-1}$}
\newcommand{\kms}{km~s$^{-1}$}
\newcommand{\kmsm}{km~s$^{-1}$~Mpc$^{-1}$}
\newcommand{\Ha}{$\rm H\alpha$}
\newcommand{\Hb}{$\rm H\beta$}
\newcommand{\hi}{{H{\sc i}}}
\newcommand{\htwo}{H$_2$}
\newcommand{\hii}{{H{\sc ii}}}
\newcommand{\nii}{\ion{N}{2}}
\newcommand{\rband}{{\em r}-band}
\newcommand{\iband}{{\em I}-band}
\newcommand{\zband}{{\em z}-band}
\newcommand{\rd}{$r_{\rm d}$}
\newcommand{\whi}{$W_{50}$}
\newcommand{\ds}{$\Delta s$}
\newcommand{\x}{$\times$}
\newcommand{\about}{$\sim$}
\newcommand{\Msun}{M$_\odot$}
\newcommand{\Lsun}{L$_\odot$}
\newcommand{\Mhi}{$M_{\rm HI}$}
\newcommand{\Fhi}{$F_{\rm HI}$}
\newcommand{\Mgas}{$M_{\rm gas}$}
\newcommand{\Mhtwo}{$M_{\rm H2}$}
\newcommand{\Mst}{$M_\star$}
\newcommand{\rmol}{$R_{\rm mol}$}
\newcommand{\Mh}{$M_{\rm h}$}
\newcommand{\must}{$\mu_\star$}
\newcommand{\nuvr}{NUV$-r$}
\newcommand{\Rinz}{$R_{50,z}$}
\newcommand{\Ropt}{$R_{\rm opt}$}
\newcommand{\sov}{$S_{0.5}$}
\newcommand{\vrot}{$V_{\rm rot}$}
\newcommand{\vs}{$V_{\rm rot}/\sigma$}
\newcommand{\cindx}{$R_{90}/R_{50}$}
\newcommand{\rhalf}{$R_{50}$}
\newcommand{\tmax}{$T_{\rm max}$}
\newcommand{\ngal}{$N_{\rm gal}$}
\newcommand{\ntot}{$N_{\rm tot}$}
\newcommand{\detfr}{$N_{\rm det}$/$N_{\rm tot}$}
\newcommand{\gi}{$g-i$}

\newcommand{\faa}{$f_{\rm AA}$}
\newcommand{\fhiar}{$f_{\rm S05}$}
\newcommand{\naa}{$N_{\rm AA}$}
\newcommand{\nhiar}{$N_{\rm S05}$}
\newcommand{\naab}{$N'_{\rm AA}$}
\newcommand{\nhiarb}{$N'_{\rm S05}$}
\newcommand{\ngass}{$N_{\rm G}$}
\newcommand{\ngassnorich}{$N_{\rm G,<AA}$}

\newcommand{\phm}{\phantom{$-$}}

\newcommand{\bct}{\vspace{-8 pt} \noindent}

\newcommand{\bc}{}

\maketitle

\label{firstpage}

\begin{abstract}

We present the extended GALEX Arecibo SDSS Survey (xGASS), a gas fraction-limited census of the
atomic {\bc hydrogen (\hi) gas content of 1179 galaxies selected only by stellar mass (\Mst $=10^{9}-10^{11.5}$ \Msun)
and redshift ($0.01<z<0.05$). This includes new Arecibo observations of 208 galaxies, for which
we release catalogs and \hi\ spectra.}
In addition to extending the GASS \hi\ scaling relations by one decade in stellar mass, we quantify 
total {\bc (atomic+molecular) cold gas fractions and molecular-to-atomic gas mass ratios, \rmol, for the subset of 
477 galaxies observed with the IRAM 30 m telescope.}
We find that atomic gas fractions keep increasing with decreasing stellar mass, with no sign of a plateau down
to $\log M_\star/M_\odot = 9$. Total gas reservoirs remain \hi-dominated {\bc across our full 
stellar mass range}, hence total gas fraction scaling relations closely resemble atomic ones, but 
with a scatter that strongly correlates with \rmol, especially at fixed specific star formation rate.
{\bc On average, \rmol\ weakly increases} with stellar mass and stellar surface density \must, {\bc but}
individual values vary by almost two orders of magnitude at fixed \Mst\ or \must. We show that, for 
galaxies on the star-forming sequence, variations of \rmol\ are mostly 
driven by changes of the \hi\ reservoirs, with a clear dependence on \must.
{\bc Establishing if} galaxy mass or structure plays the most important role in regulating the cold
gas content of galaxies {\bc requires an accurate separation} of bulge and disk components for the study of
gas scaling relations.

\end{abstract}

\begin{keywords}
galaxies: evolution -- galaxies: ISM -- radio lines: galaxies -- galaxies: fundamental parameters
\end{keywords}

\section{Introduction}\label{s_intro}

The gas-star formation cycle is central to the formation and evolution of galaxies
{\bc (see \eg\ \citealt{leroy08,lilly13} and review by \citealt{kennicutt12})}.
Understanding the complex interplay between the various components {\bc (such as multi-phase neutral and 
ionised gas, and dust) of the interstellar medium (ISM; \citealt{mckee77,cox05})}
and star formation
as a function of galaxy properties, environment and cosmic time is a formidable task, which
requires sensitive measurements across the electromagnetic spectrum and on multiple spatial scales
for statistical data sets, as well as detailed numerical simulations to gain insights into the 
physical processes involved.

Even when we restrict ourselves to the global properties of galaxies in the local Universe,
gathering the necessary data remains challenging. The main limitation comes from the paucity
{\bc of measurements of the cold gas content}\footnote{
{\bc By {\it cold gas} we refer to neutral hydrogen, both atomic and molecular.
\hi\ gas is typically found in two phases, a cold neutral medium (CNM, $T\lesssim 300$ K, best traced 
by \hi\ in absorption) with a cloudy structure and a diffuse, warm neutral medium 
(WNM, $T\gtrsim 5000$ K, providing the bulk of the \hi\ seen in emission; \citealt{brinks90,wolfire95,kalberla09});
molecular hydrogen is found in dense clouds with lower temperatures ($T\sim 10-20$ K;
\citealt{brinks90,klessen16}).
}}
for large, representative galaxy samples compared to optical, infrared, or
ultraviolet surveys. Blind surveys of atomic hydrogen such as the \hi\ Parkes All-Sky Survey 
\citep{hipass,hipass_cat,hipass_north},
and the Arecibo Legacy Fast ALFA survey \citep[ALFALFA;][]{alfalfa,alfalfa40} measured the \hi\ content 
for \about 50,000 galaxies, but detect only the most gas-rich systems in most of their volume
({\bc \about 7000 deg$^2$ and} $z<0.06$ for ALFALFA). Samples of molecular hydrogen content, which 
is traced by carbon monoxide ($^{12}$CO, hereafter CO)
line emission, are almost two orders of magnitude smaller
\citep{fcrao,coldgass1,boselli14a,allsmog,allsmog_dr}. As a result, accurate constraints 
for key parameters such as the molecular-to-atomic {\bc gas mass} ratio as a function of galaxy properties for 
unbiased samples are still scarce. 

It is indeed generally accepted that atomic hydrogen has to transition into molecular phase
in order to fuel star formation \citep{blitz_ros,bigiel08,leroy08,krumholz09}, although molecular gas could 
just be tracing star formation, formed as the by-product of the gravitational collapse of atomic gas \citep{glover12}.
The partition of total cold gas into \hi\ and \htwo\ and the efficiency of the atomic-to-molecular conversion 
are thus crucial quantities to measure in order to determine the physical processes regulating 
the star formation cycle in galaxies.

Substantial observing effort in the past decade went into measuring atomic and molecular gas masses 
for large samples of galaxies selected from optical surveys, and largely missed by \hi -blind surveys 
\citep[\eg,][]{atlas3d_hi,atlas3d_co,boselli14a}.

Our GALEX Arecibo SDSS Survey \citep[GASS;][]{gass1} was designed to investigate the main gas fraction 
scaling relations for a representative {\bc (in terms of \hi\ content)}, {\it stellar mass-selected} sample 
of galaxies with stellar masses greater than $10^{10}$ \Msun. The gas fraction limited nature of our 
observations means that integration times 
on each source were dictated by the request to reach gas fraction limits of \about 2\%, thus providing the
most sensitive \hi\ measurements for a large sample currently available.
The combination of GASS on Arecibo and its follow-up program on the IRAM 30m telescope 
\citep[COLD GASS survey,][]{coldgass1} resulted in a benchmark multi-wavelength data set, including physical 
information about the stars and both atomic and molecular hydrogen gas phases in massive systems. 

There were very good reasons to extend GASS and COLD GASS down to a stellar mass of $10^9$ \Msun. 
First, these extensions would allow us to probe a crucial ``sweet spot'' for understanding the physical
processes that regulate the conversion of gas into stars and shape star-forming galaxies, without 
the additional complexities introduced by the presence of massive bulges and active galactic nuclei 
that are ubiquitous in the GASS stellar mass regime. 
Second, the {\bc scatter in} the gas fraction scaling relations is expected to be driven by intrinsic 
properties of the disks, such as amount of angular momentum and stellar surface density \citep[\eg,][]{fu10}. 
Again, testing this prediction with GASS is hampered by the presence of massive
bulges, which could influence gas content as well \citep[\eg,][]{martig09}. 
The new observations target a stellar mass regime that is dominated by star-forming disks, thus 
greatly alleviating these limitations. As showed by GASS, examining the scatter around the mean relations, and
particularly its second-parameter dependencies, requires statistical samples of several hundred galaxies.

Here we present the complete low-mass extension of GASS, hereafter {\it GASS-low},
which includes new Arecibo observations of 208 galaxies.
The combination of GASS and GASS-low, which we refer to as the {\it extended GASS} (xGASS) survey, results
in a representative sample of 1179 galaxies covering the $9<$ log \Mst/\Msun $<11.5$ 
stellar mass interval (see Section~\ref{s_repr}).

The companion extension of the molecular gas survey, {\it COLD GASS-low}, and the properties
of the full xCOLD GASS sample are presented in \citet{xcoldgass}.
Unlike the original GASS and COLD GASS surveys that were designed to explore
the transition between star-forming and passive galaxies, these low-mass extensions
aim to understand the basic physical processes governing star-forming galaxies.

This paper is organized as follows. In Section~\ref{s_xgass} we describe the sample selection and Arecibo 
observations of GASS-low galaxies, and combine these with GASS to obtain the xGASS representative sample.
This includes the correct proportion of \hi -rich ALFALFA galaxies that were not targeted to increase
survey efficiency, and thus is representative of the \hi\ properties of the galaxy population in our
stellar mass and redshift intervals. We summarize in Section~\ref{s_xcoldgass} the main properties of
the xCOLD GASS survey, and discuss the overlap sample with both \hi\ and \htwo\ observations.
Our \hi, total gas, and \htwo/\hi\ scaling relations are presented in Section~\ref{s_results};
model comparisons are discussed in Section~\ref{s_models}. We summarize our main findings and 
conclude in Section~\ref{s_concl}. The GASS-low data release can be found in Appendix~A.

All the distance-dependent quantities in this work are computed assuming a cosmology with 
$H_0 = 70$ \kmsm, $\Omega_m=0.3$ and $\Omega_\lambda=0.7$. We assume a \citet{chabrier03} initial mass 
function. AB magnitudes are used throughout the paper.

\section{\lowercase{x}GASS: the extended GASS survey}\label{s_xgass}

\subsection{The low-mass extension of GASS}

\subsubsection{Sample selection and survey strategy}\label{s_sample}

The galaxies of the GASS low-mass extension were {\bc selected from a parent sample
of 872 sources} extracted from the intersection of the SDSS DR7 \citep{sdss7} spectroscopic survey, 
the GALEX Medium Imaging Survey \citep{galex} and projected ALFALFA footprints according to 
the following criteria:
\begin{itemize}
\item Stellar mass $9.0 < {\rm log} M_\star/M_\odot < 10.2$
\item Redshift $0.01 < z < 0.02$.

\end{itemize}

{\bc Because GASS-low targets smaller galaxies than GASS, we lowered the redshift interval
to ease their detection (GASS was limited to $0.025 < z < 0.05$).}
Galaxies in these stellar mass and redshift intervals have angular diameters smaller than 1 arcmin (as in GASS). 
Thus our targets fit comfortably within a single SDSS frame and GALEX pointing, so that accurate photometry 
(and hence stellar masses and star formation rates) can be measured, and a single pointing of the IRAM 30m 
telescope recovers an accurate total CO flux in most cases.

{\bc For GASS, we imposed a flat stellar mass distribution for our targets, in order to ensure enough statistics
at the high-mass end. Similarly, we sampled the stellar mass interval of GASS-low galaxies roughly evenly 
(see Section~\ref{s_weights}).}

In order to optimize survey efficiency, we prioritized the observations of the galaxies lying within
the ALFALFA 40\% \citep[hereafter AA40;][]{alfalfa40} footprint and/or galaxies already observed
with the IRAM telescope. Galaxies with good quality \hi\ detections already available from AA40 or the Cornell 
\hi\ digital archive \citep[][hereafter S05]{springob05} were not reobserved (see Section~\ref{s_repr} below).

Following the GASS strategy, we observed the targets until detected, or until a limit of a few percent in gas mass fraction 
(\Mhi/\Mst) was reached. Practically, we set a limit of $M_{\rm HI}/M_\star > 0.02$ for galaxies with 
${\rm log} (M_\star/M_\odot) >9.7$, and a constant gas mass limit ${\rm log} (M_{\rm HI}/M_\odot) =8$ for galaxies with 
smaller stellar masses. This corresponds to a gas fraction limit of $0.02-0.1$ for the whole sample.

Given the \hi\ mass limit assigned to each galaxy (set by its gas fraction limit and stellar mass), we computed 
the on-source observing time, \tmax, required to reach that value with our observing mode and instrumental setup,
assuming a velocity width of 200 \kms, smoothing to half width, and signal-to-noise of 5. The \tmax\
values thus obtained vary between 1 and 95 minutes.

\subsubsection{Arecibo observations and data reduction}

GASS-low observations started in August 2012 and ended in May 2015. These were scheduled in 100 
observing runs under Arecibo programs A2703 and A2749; the total telescope time allocation was 
263 hours.

The observing mode and data reduction were the same as GASS. 
All the observations were carried out remotely in standard
position-switching mode, using the L-band wide receiver and the
interim correlator as a backend. Two correlator boards with 12.5 MHz
bandwidth, one polarization, and 2048 channels per spectrum (yielding
a velocity resolution of 1.4 \kms\ at 1370 MHz before smoothing) were centered
at or near the frequency corresponding to the SDSS redshift
of the target. We recorded the spectra every second with 9-level sampling.

The data reduction, performed in the IDL environment, includes
Hanning smoothing, bandpass subtraction, excision of radio frequency interference (RFI), and flux
calibration. The spectra obtained from each on/off pair are weighted by 1/$rms^2$,
where $rms$ is the root mean square noise measured in the
signal-free portion of the spectrum, and co-added. The two
orthogonal linear polarizations (kept separated up to this point) are
averaged to produce the final spectrum; {\bc polarization mismatch, if significant, is noted 
in Appendix~B. The spectrum is then} boxcar smoothed,
baseline subtracted and measured as explained in \citet{gass1}. 
The instrumental broadening correction for the velocity widths
is described in \citet[see also \citealt{gass4}]{gass_dr2}.
Our RFI excision technique is illustrated in detail in \citet{highz}.

\subsubsection{The new data release}\label{s_data}

This data release includes new Arecibo observations of 208 galaxies. The catalogs of optical, UV
and 21 cm parameters for these objects are presented in Appendix~A.

All the optical parameters were obtained from the SDSS DR7 database server\footnote{
{\em http://cas.sdss.org/dr7/en/tools/search/sql.asp}
}. Stellar masses are from the Max Planck Institute for Astrophysics (MPA)/Johns Hopkins
University (JHU) value-added catalog based on SDSS DR7\footnote{
{\em http://www.mpa-garching.mpg.de/SDSS/DR7/}; we used the improved stellar masses from
{\em http://home.strw.leidenuniv.nl/\about jarle/SDSS/}
}, and assume a \citet{chabrier03} initial mass function.

UV photometry and star formation rate (SFR) measurements were obtained for
the full xGASS sample as explained in detail by \citet{steven1}.
Briefly, NUV magnitudes are typically from the {\it GALEX Unique Source Catalogs}\footnote{
{\em http://archive.stsci.edu/prepds/gcat/}
} \citep{seibert12}, or other GALEX catalogs such as BCSCAT \citep{bcscat} and the 
GR6+7 data release\footnote{
{\em http://galex.stsci.edu/GR6/}
}. The measured \nuvr\ colors are corrected for Galactic
extinction following \citet{wyder07}, from which we obtained
$A_{NUV}-A_r = 1.9807 A_r$ (where the extinction $A_r$ is available
from the SDSS data base and reported in Table~\ref{t_sdss} below). We
did not apply internal dust attenuation corrections.

SFRs were computed combining NUV with mid-infrared (MIR) fluxes from the Wide-field Infrared 
Survey Explorer \citep[WISE,][]{wise}. We performed our own aperture photometry on the 
WISE atlas images and used the w4 band (22 \micron) measurements when possible, and
w3 band (12 \micron) ones otherwise. If good NUV and MIR fluxes were both not available, 
we used SFRs determined from the spectral energy distribution fits of \citet{gass3};
{\bc we refer the reader to \citet{steven1} for further details.}

\begin{figure*}
\begin{center}
\includegraphics[width=14cm]{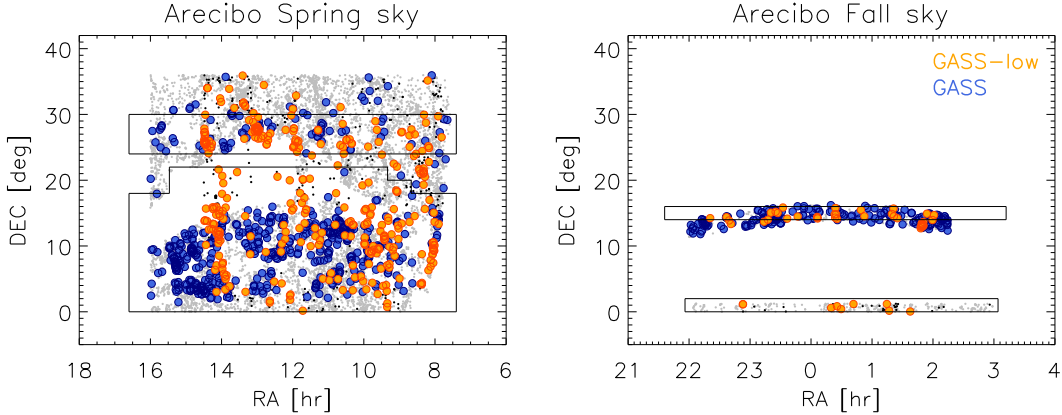}
\caption{Sky distribution of galaxies in the GASS (blue) and GASS-low (orange) representative samples.
Black and gray dots show the parent samples of the two surveys (GASS-low and GASS, respectively).
The areas enclosed by thin black lines indicate the ALFALFA 70\% footprint in the regions of interest.
}
\label{sky}
\end{center}
\end{figure*}

The catalogs presented in our three GASS data releases \citep{gass1,gass_dr2,gass_dr3}
and in this work, as well as the full xGASS representative sample (see Section~\ref{s_repr}), 
are available on the xGASS website\footnote{
{\em http://xgass.icrar.org}
}, along with all the Arecibo \hi\ spectra in digital format.

\subsection{The xGASS representative sample}\label{s_repr}


In order to increase survey efficiency we did not reobserve galaxies
with good quality \hi\ detections {\bc in ALFALFA (based on the most recent data release
available at the time of our observations, which was AA40 for GASS-low)}
or the Cornell \hi\ digital archive \citep[][hereafter S05]{springob05}. For ALFALFA,
this refers to galaxies with detection code ``1'' (\ie, signal-to-noise 
SNR $\geq 6.5$); sources identified by code ``2'' (with lower SNR  
but coincident with an optical counterpart at the same redshift) were reobserved.
Hence both GASS and GASS-low {\it observed} samples lack \hi -rich objects, which 
must be added back in the correct proportions to obtain data sets that are
{\it representative} in terms of \hi\ properties. Because the two surveys
cover different volumes and stellar mass regimes, we generate the two
representative samples separately, {\bc taking advantage in both cases of the more recent 
70\% data release\footnote{
Obtained from {\it http://egg.astro.cornell.edu/alfalfa/data/index.php}
}
of ALFALFA (AA70)}. This is done slightly differently to the three 
GASS data releases.
We explain below the procedure used to generate the xGASS representative sample, which 
is simply obtained by joining the GASS-low and (revised) GASS ones. 

First, we divide each sample into two parts: inside and outside the
{\bc AA70} footprint (see Fig.~\ref{sky}), which is given by the sky distribution 
of the 23,881 {\bc \hi-}detected galaxies included in the publicly available catalog.
The fraction of GASS-low and GASS parent samples included by this footprint are
74\% and 75\%, respectively.
We compute the ALFALFA detection fraction in each stellar mass bin, \faa,
defined as the ratio of number of galaxies \naa\ detected by ALFALFA (code ``1'' only) and
{\bc total} number of sources $N_{\rm PS}$ in the parent sample, both restricted to the sky region with complete
ALFALFA coverage and to the given stellar mass bin. Detection fractions decrease
from 56.5\% to 50.4\% for GASS-low and from 16.2\% to 13.9\% for GASS going from the
lowest to the highest stellar mass bin.

Second, we generate the representative sample for the subset within the {\bc AA70} footprint
by adding the correct proportion of gas-rich, ALFALFA detections in each stellar mass bin.
If $N_{\rm poor}$ is the number of observed galaxies in the given stellar mass bin that
are not detected by ALFALFA, we obtain the number of gas-rich galaxies to be added as follows:
\begin{equation}
N_{\rm rich} = N_{\rm poor} \times \frac{N_{\rm AA}/N_{\rm PS}}{1-N_{\rm AA}/N_{\rm PS}} = N_{\rm poor} \times \frac{f_{\rm AA}}{1- f_{\rm AA}}
\end{equation}
where $N_{\rm PS}$ is the number of galaxies in the parent sample within the {\bc AA70} footprint
in the given stellar mass bin. We denote as $N_{\rm xGASS,rich}$ those galaxies that we observed 
but are also ALFALFA detections (for instance, GASS galaxies outside the AA40 footprint 
that turned out to be detected in {\bc AA70}). These galaxies are not included in $N_{\rm poor}$, 
but we pick them first as gas-rich systems to be added to the sample. Next, we select 
$N_{\rm rich} - N_{\rm xGASS,rich}$ {\bc uniformly distributed, random} galaxies from ALFALFA (giving first preference to sources with
xCOLD GASS data) and we add them to $N_{\rm poor} + N_{\rm xGASS,rich}$ to obtain our representative sample.

\begin{figure*}
\begin{center}
\includegraphics[width=14cm]{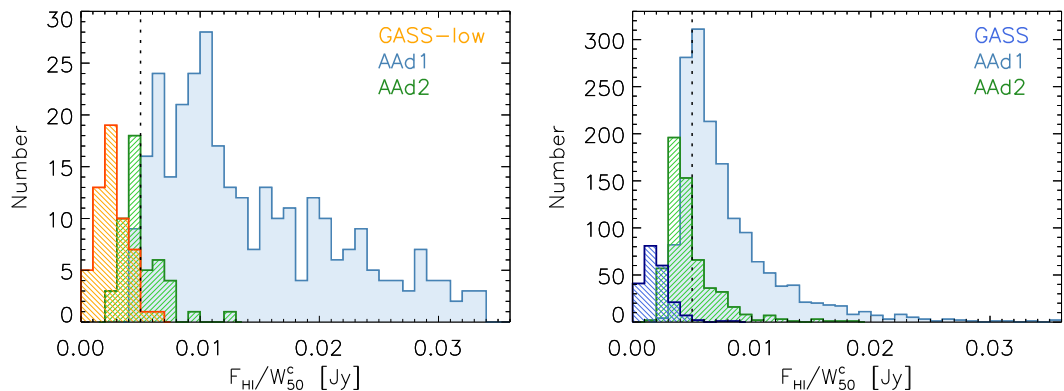}
\caption{Histograms of the mean flux density across the detected \hi\ signal for GASS-low (left) and
GASS (right) surveys, with ALFALFA detections (quality codes 1 and 2) in the corresponding volumes.
Dotted lines indicate the value adopted to separate gas-rich from gas-poor galaxies in regions
of sky not covered by ALFALFA (see text).
}
\label{AAdetlim}
\end{center}
\end{figure*}

Third, we deal with the part of the sample outside the {\bc AA70} footprint. Here we need to estimate which 
galaxies would be detected by ALFALFA if the survey was complete\footnote{
{\bc This step is necessary for GASS-low due to the inclusion of S05 galaxies, and we wanted to treat these and
our Arecibo observations in a uniform way. Furthermore, GASS lacks \hi-rich galaxies
at high stellar masses outside the AA70 footprint, most likely due to a combination of 
large-scale-structure and Arecibo time allocation, so our procedure corrects for this. However, as already
noted, both GASS and GASS-low samples are dominated by the subsets within the AA70 footprint.}
}. Because the sensitivity (or completeness) of 
ALFALFA depends on both flux and velocity width {\bc of the \hi\ signal
\citep[see Section 6 of][]{alfalfa40}, we inspect the histogram of the \hi\ integrated flux, {\bc \Fhi,} divided 
by the observed velocity width, $W_{50}^c$ (measured at the 50\% peak level and corrected for instrumental broadening 
and redshift only, see Appendix~A)} to decide where to set the threshold.
Fig.~\ref{AAdetlim} shows the {$\bc F_{\rm HI}/W_{50}^c$} histograms for GASS-low (left) and GASS (right) galaxies,
compared with ALFALFA detections within the corresponding parent samples (code ``1'' and ``2'' are 
indicated in light blue and green, respectively). For both surveys we adopt a value of 0.005 Jy
as our threshold (dotted lines), below which ALFALFA code ``2'' sources start dominating over
high signal-to-noise ones. We verified that changing this number slightly does not have a significant
effect on the final sample (changing the sky footprint over which detection fractions are computed, 
e.g. from {\bc AA40 to AA70}, has a much larger effect).
Then we use equation~1 to generate our representative samples, where $N_{\rm poor}$ now
includes xGASS galaxies with {$\bc F_{\rm HI}/W_{50}^c<0.005$} Jy (\ie, below the ALFALFA detection limit) and
$N_{\rm xGASS,rich}$ those above this threshold. The \hi-rich galaxies are extracted randomly from
ALFALFA detections not already in the sample, trying to maximize the overlap with xCOLD GASS.

In our GASS papers, we treated the S05 \hi\ archive in a similar way as ALFALFA: we computed the fraction
of parent sample galaxies with \hi\ data in the archive, \fhiar\ (not including ALFALFA detections), 
and used equation~1 (with \fhiar\ replacing \faa) to obtain the number of \hi-rich S05 galaxies to be 
added to the observed sample. While this does not affect our scaling relations (only 1.3\% of the galaxies in the
GASS DR3 representative sample were from S05\footnote{
Contrary to what we did for GASS in our previous papers, we now use ALFALFA instead of S05 \hi\ fluxes for galaxies detected
in both catalogs. While S05 integration times are typically longer, the spectra were obtained with a variety 
of single-dish radio telescopes, hence have variable sensitivity, spatial and spectral resolutions. Thus, 36 out of
760 galaxies in the GASS DR3 representative sample had \hi\ measurements from S05, but only 10 of these are 
not detected by ALFALFA.}
), this is not entirely correct, because the S05 archive is 
not an \hi-blind survey, and thus \fhiar\ is not a meaningful detection fraction. Thus we no longer add
\hi-rich S05 galaxies to the observed sample.

We also considered including S05 galaxies {\it below} the ALFALFA detection threshold, together with the right
complement of ALFALFA detections, to increase our statistics. This cannot be done in the GASS volume,
where we would add 104 ``\hi-poor'' S05 galaxies, because the \hi\ archive sample is deeper than ALFALFA 
but still \hi-rich compared to GASS (thus we would bias our sample). However, this is not the case for the 
GASS-low volume, where there are only 13 ``\hi-poor'' S05 galaxies, all with gas fractions comparable to
our observations. Thus we include these in our sample as if they had been observed by us (\ie\ increasing 
$N_{\rm poor}$ in equation~1), and verified that our scaling relations are not affected by this choice.

\begin{figure*}
\begin{center}
\includegraphics[width=16cm]{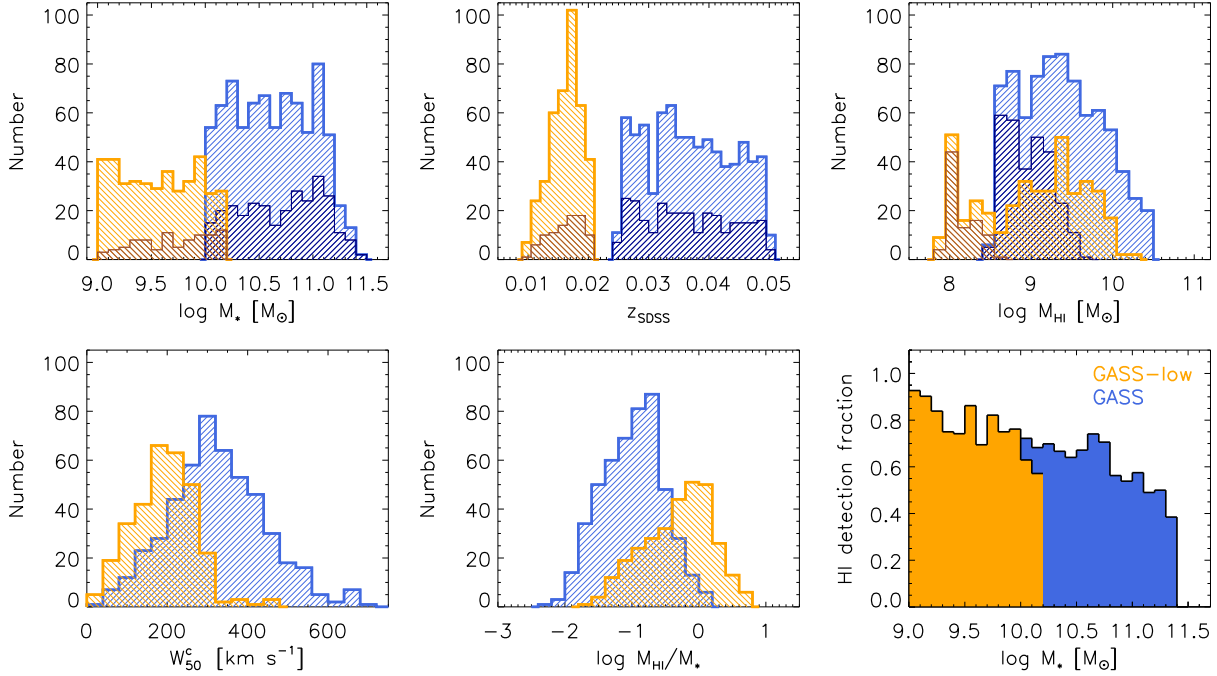}
\caption{xGASS {\bc representative} sample properties. {\it Top row:} Distributions of stellar mass, optical redshift and \hi\ mass
for GASS (blue) and GASS-low (orange); {\bc dark blue and brown histograms indicate \hi\ non-detections
for the two surveys, respectively.}
{\it Bottom row:} Distributions of \hi\ velocity width {\bc (corrected for instrumental broadening 
and redshift) and \hi\ gas fraction for detected galaxies; the \hi\ detection fraction 
(\ie, the ratio of detections to total) as a function of stellar mass is shown on the right panel} (blue: GASS; orange: GASS-low).
}
\label{hist}
\end{center}
\end{figure*}

The new GASS and GASS-low representative samples include 781 and 398 galaxies respectively, 
for a total 1179 xGASS galaxies. Fig.~\ref{hist} summarizes the main properties of the sample. On the top row,
the distributions of stellar mass, optical redshift from SDSS and \hi\ mass are shown separately for 
GASS (blue) and GASS-low (orange); non-detections are indicated in dark blue and {\bc brown}, respectively. On the
bottom row, we show the histograms of velocity width and gas fraction for the \hi\ detections, as well as
the \hi\ detection fraction as a function of stellar mass for the two surveys. {\bc The \about 10\% detection rate
difference in the two overlapping stellar mass bins is most likely just noise
(a couple more detections in GASS-low would have brought the detection rates into agreement).}
The observed velocity widths 
peak at \about 200 \kms\ and 300 \kms\ for GASS-low and GASS, respectively, so these are the values that
we adopt to compute upper limits for the \hi\ mass of non-detections in the two surveys.

\subsubsection{Recovering a volume-limited sample}\label{s_weights}

In our previous GASS work, we computed average scaling relations by weighting each
measured gas fraction (detection or upper limit) by a factor $w_i(\rm M_\star)$, in order to compensate for 
the flat stellar mass distribution imposed on the survey. Weights were computed using 
the parent sample as a reference, by binning both parent and representative samples by 
stellar mass and taking the ratio between the two histograms.

\begin{figure}
\begin{center}
\includegraphics[width=7cm]{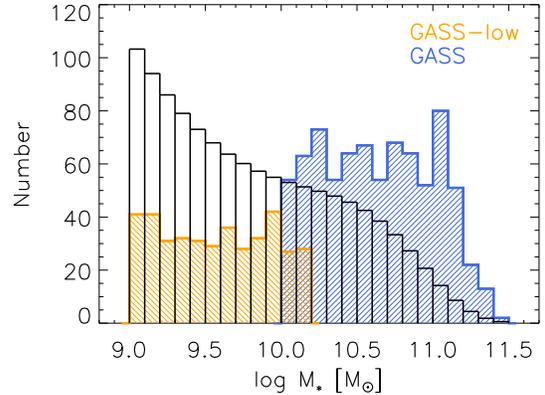}
\caption{Stellar mass distributions for the xGASS representative sample (orange and blue indicate
GASS-low and GASS subsets, respectively) and for a volume-limited sample with the same number of 
galaxies (see text).
}
\label{weights}
\end{center}
\end{figure}

The xGASS representative sample has a similar problem, largely due to the difference 
in sample size between GASS and its low mass extension -- low mass galaxies are 
under-represented and high mass ones over-represented, compared to what is expected 
for a volume-limited sample. This is illustrated in Figure~\ref{weights}, 
which shows the stellar mass distributions for GASS-low (orange histogram) and GASS (blue)
representative samples, and for a volume-limited sample with the same total number of
galaxies (black). The latter was obtained by sampling the local stellar mass function,
parameterized as a double Schechter function\footnote{
We use the logarithmic form from \citet{moffett16}.
} by \citet{baldry12}:
\begin{displaymath}
  \Phi(\log M) d\log M = \ln(10) ~e^{-x} 
  \left[ \phi_1^* ~x^{(1+\alpha_1)}+\phi_2^* ~x^{(1+\alpha_2)} \right] d\log M 
\end{displaymath}

\noindent
where $x =10^{\log M -\log M^*}$, $\log M^* =10.66$~\Msun, $\phi_1^* =0.00396$, $\phi_2^* =0.00079$, 
$\alpha_1 =-0.35$ and $\alpha_2 =-1.47$.

Thus, in this work we use the above stellar mass function to weight the gas fraction measurements
when we compute average and median gas scaling relations; weights are simply obtained as the ratio between 
the black and the colored histograms in Figure~\ref{weights} for the corresponding stellar mass bin and 
survey (GASS or GASS-low).

\begin{figure*}
\begin{center}
\includegraphics[width=14cm]{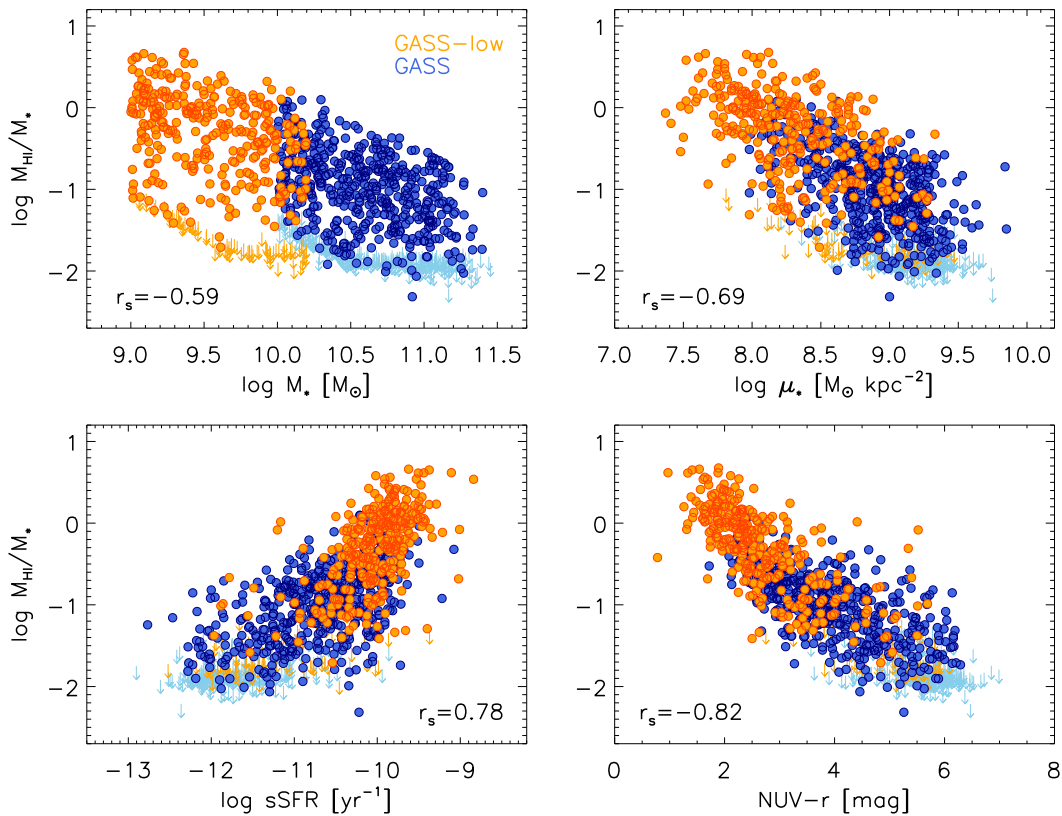}
\caption{xGASS scaling relations: the \hi\ mass fraction is plotted as a function of stellar
mass, stellar mass surface density, specific SFR and observed \nuvr\ color. Circles and downward arrows
indicate individual \hi\ detections and {\bc 5$\sigma$ upper limits}, respectively, with the new GASS-low observations {\bc shown 
in orange. The Spearman's correlation coefficient, $r_s$, is given on the bottom of each panel.}
}
\label{scalings}
\end{center}
\end{figure*}

\section{\lowercase{x}COLD GASS: The molecular gas survey}\label{s_xcoldgass}

The COLD GASS survey measured homogeneous molecular gas masses, via the CO(1-0) emission line
fluxes, for 366 galaxies extracted from the GASS sample \citep{coldgass1}. Because the IRAM
beam size at the frequency of the observed CO(1-0) transition is 22\arcsec, aperture corrections
were applied to extrapolate the measured CO line fluxes to total fluxes, as described in 
\citet{saintonge12}. The extension of COLD GASS to a stellar mass of $10^9$ \Msun, {\it COLD GASS-low},
includes IRAM observations of 166 additional galaxies, {\bc randomly extracted from the GASS-low parent 
sample (see Section~\ref{s_sample})}. The two surveys taken together constitute
{\it xCOLD GASS}, which includes 532 galaxies (333 detections) and is described in detail in \citet{xcoldgass}.

The CO(1-0) fluxes are converted into \htwo\ molecular masses using a multivariate conversion function, $\alpha_{CO}$, 
following \citet{accurso17}. This function depends primarily on metallicity and secondarily on the offset
from the star-forming main sequence, \ie\ a parameter related to the strength of the UV radiation field; $\alpha_{CO}$ values 
for xCOLD GASS detections vary between 1 and 24.5 \Msun~(K~\kms~pc$^2$)$^{-1}$, with an average of 4.44
(for comparison, the Galactic value is 4.35), taking into account the contribution of Helium. In this work, we
use molecular gas masses without Helium contribution, \Mhtwo, to compute molecular-to-atomic hydrogen gas 
mass ratios.

The overlap between xCOLD GASS and the xGASS representative sample, which we refer to as {\it xGASS-CO},
includes 477 galaxies (290 CO detections) and is used in this work to investigate total gas scaling relations and \htwo/\hi\ mass ratios.
The remaining 55 galaxies with CO data are not included in xGASS because of one of the following reasons:
(a) lack of \hi\ observations (13); (b) specifically targeted by COLD GASS for their very high specific SFRs,
hence not preferentially selected for our representative sample (35, 2 of which were
randomly picked as ALFALFA ``code 1'' sources); (c) S05 detections in GASS (7); or (d) ALFALFA ``code 1'' 
sources that were not selected because the stellar mass bin already included enough \hi-rich systems (2).

We recomputed the weights for xGASS-CO in order to recover the stellar mass distribution of 
a volume-limited sample, following the procedure described in Section~\ref{s_weights}. This 
sample is representative in terms of \hi\ content (we verified that the average \hi\ scaling relations 
obtained for xGASS and for xGASS-CO are consistent within the errors).

\section{Results}\label{s_results}

We briefly revisit the main \hi\ gas fraction scaling relations, which
extend our previous work \citep{gass1,gass_dr2,gass_dr3} to lower stellar masses,
and take advantage of the combined \hi\ and \htwo\ data set to investigate total
gas scaling relations. Molecular gas scaling relations are presented in a 
companion paper \citep{xcoldgass}. As discussed below, the distinct
behavior of the atomic and molecular phases at low stellar masses motivates
a more detailed discussion of the molecular-to-atomic gas {\bc mass} ratio of galaxies
along and outside the star-forming sequence.

\begin{figure*}
\begin{center}
\includegraphics[width=14cm]{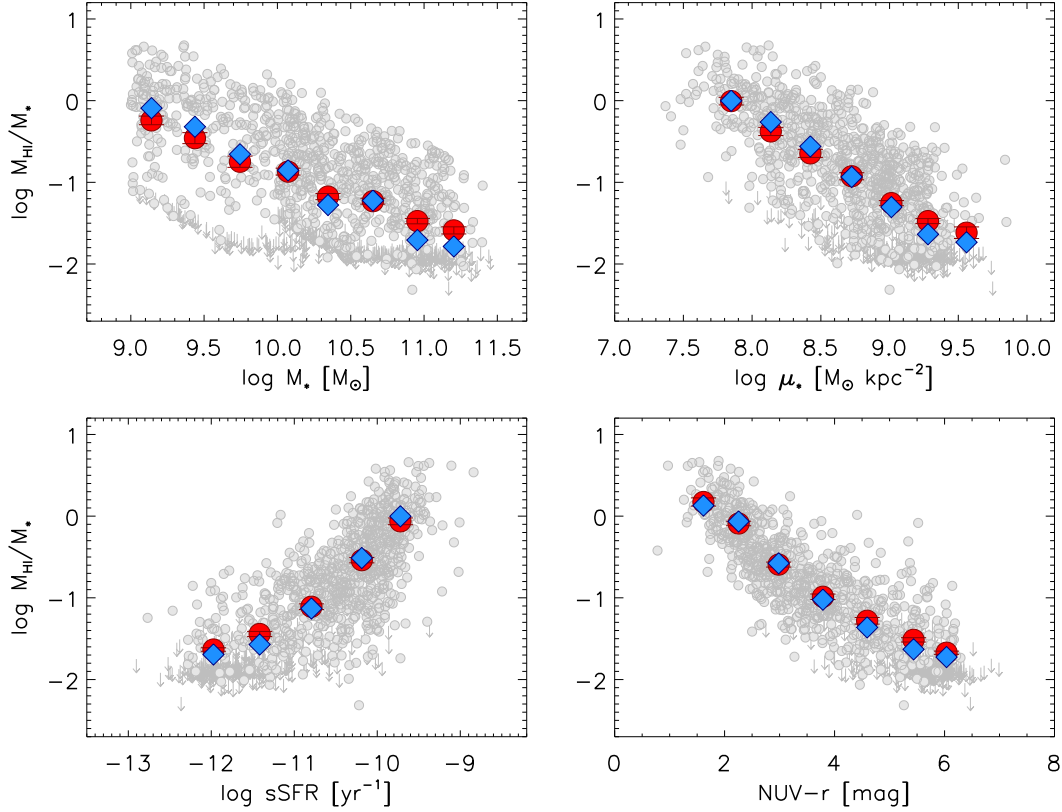}
\caption{Average trends of \hi\ mass fraction as a function of stellar mass, stellar mass surface density, 
specific SFR and observed \nuvr\ color for the xGASS sample. In each panel, large red circles and blue diamonds 
indicate weighted averages and weighted medians of the logarithms of the gas fractions, respectively (see
text). Only bins including at least 20 galaxies are shown. Error bars are errors on the weighted means.
These results are listed in Table~\ref{t_avgs}. Small circles and downward arrows show individual \hi\ detections 
and non-detections (plotted at their upper limits), respectively.
}
\label{scalings_avgs}
\end{center}
\end{figure*}

\subsection{Atomic gas fraction scaling relations}

The \hi\ gas fraction scaling relations are presented in Figure~\ref{scalings}.
Clockwise from the top left, we show how the gas mass fraction \Mhi/\Mst\ varies with stellar mass, 
stellar mass surface density, observed \nuvr\ color and specific SFR (sSFR) for the full xGASS 
representative sample. Circles and downward arrows indicate \hi\ detections 
and \hi\ upper limits, respectively, with {\bc the new GASS-low survey shown in orange}.
The low mass galaxies smoothly extend GASS trends by one dex in stellar mass, probing
higher gas fractions and sSFRs, bluer colors and lower stellar surface densities
typical of disk-dominated systems. {\bc Because some of these relations do not appear
linear (especially those with color and sSFR), we quantified their strength with the 
Spearman's rank correlation coefficient, $r_s$, computed including the upper limits.
The most significant correlation is with \nuvr\ color ($r_s=-0.82$), with the \Mst\ and
\must\ relations having significantly lower correlation coefficients ($r_s=-0.59$ and $-0.69$, respectively).}

This is better seen in Figure~\ref{scalings_avgs}, which quantifies the observed trends
in terms of average (large red circles) and median (blue diamonds) gas fractions. The values
plotted are weighted means and weighted medians\footnote{
Given n elements $x_1 ... x_n$ with positive weights $w_1 ... w_n$ such that
their sum is 1, the weighted median is defined as the element $x_k$ for which:
$\displaystyle \sum_{x_i < x_k} w_i < 1/2$ and 
$\displaystyle \sum_{x_i > x_k} w_i \leq 1/2$.
}
of the logarithm of the gas fractions, where the weighting is applied to correct for the
stellar mass bias of the sample (see Section~\ref{s_weights}); non-detections were set to their
upper limits. For reference, small gray circles and downward arrows reproduce individual \hi\ 
detections and upper limits from Figure~\ref{scalings}. Average and median gas fractions
track each other closely in all plots, despite the fact that the underlying distributions
are clearly not log-normal; however medians are preferable descriptors because less
sensitive to the treatment of non-detections, which could lie anywhere below
the upper limits. The values of the weighted average and median gas fractions shown in 
this figure are listed in Table~\ref{t_avgs}.
In order to quantify the scatter in these relations, we computed the difference between
the 75th and 25th percentiles of the $\log$(\Mhi/\Mst) distributions in each bin ($\Delta_{75-25,i}$, {\bc including 
the \hi\ non-detections at their upper limits}), 
and took their {\bc arithmetic mean} ($\bar{\Delta}$). We obtained $\bar{\Delta}=0.96, 0.73, 0.51$ and 0.43 dex for the
\Mst, \must, sSFR and \nuvr\ relations, respectively, with no clear trends in $\Delta_{75-25,i}$
with any of the above quantities (except for an artificially lower scatter in the two bins dominated by
non-detections in each relation).

Figures~\ref{scalings} and \ref{scalings_avgs} show that
\hi\ gas fraction is more tightly related to \nuvr\ color and sSFR, and both relations
steepen in the star-forming sequence (approximately corresponding to \nuvr $<3$ mag and $\rm \log sSFR~[yr^{-1}] >-10.5$). 
This change of slope could be due to a saturation effect at the opposite end, where
we hit the survey sensitivity limit and upper limits dominate the statistics.
Contrary to the molecular gas fraction, which correlates more strongly with sSFR \citep{xcoldgass}, 
the atomic gas fraction is more tightly related to \nuvr\ color ($\bar{\Delta}=0.43$ dex), which traces
dust-unobscured star formation \citep{bigiel_m83}.

\begin{table}
\centering
\caption{\hi\ gas fraction scaling relations for xGASS}
\label{t_avgs}
\begin{tabular}{lrrcr}
\hline\hline
$x$  & $\langle x \rangle$ & $\langle M_{\rm HI}/M_\star \rangle^{a}$  & $(M_{\rm HI}/M_\star)^{b}$   &  $N^{c}$ \\
\hline
log \Mst  &    9.14 &  $-$0.242$\pm$0.053 &  $-$0.092 &    113 \\
 	  &    9.44 &  $-$0.459$\pm$0.067 &  $-$0.320 &     92 \\
  	  &    9.74 &  $-$0.748$\pm$0.069 &  $-$0.656 &     96 \\
 	  &   10.07 &  $-$0.869$\pm$0.042 &  $-$0.854 &    214 \\
  	  &   10.34 &  $-$1.175$\pm$0.037 &  $-$1.278 &    191 \\
  	  &   10.65 &  $-$1.231$\pm$0.036 &  $-$1.223 &    189 \\
  	  &   10.95 &  $-$1.475$\pm$0.033 &  $-$1.707 &    196 \\
  	  &   11.20 &  $-$1.589$\pm$0.044 &  $-$1.785 &     86 \\
          &	    &  		          &	      &        \\ 
log \must &    7.85 &  $-$0.006$\pm$0.047 &  $-$0.002 &     61 \\
 	  &    8.14 &  $-$0.377$\pm$0.050 &  $-$0.262 &    129 \\
 	  &    8.42 &  $-$0.646$\pm$0.049 &  $-$0.561 &    160 \\
 	  &    8.72 &  $-$0.926$\pm$0.044 &  $-$0.934 &    221 \\
 	  &    9.01 &  $-$1.255$\pm$0.032 &  $-$1.303 &    326 \\
 	  &    9.28 &  $-$1.475$\pm$0.031 &  $-$1.636 &    233 \\
 	  &    9.56 &  $-$1.617$\pm$0.071 &  $-$1.734 &     30 \\
          &	    &  		          &	      &        \\ 
log sSFR  &  $-$11.97 & $-$1.633$\pm$0.022 &  $-$1.694 &    204 \\
 	  &  $-$11.42 & $-$1.442$\pm$0.032 &  $-$1.571 &    214 \\
 	  &  $-$10.79 & $-$1.109$\pm$0.034 &  $-$1.130 &    233 \\
 	  &  $-$10.19 & $-$0.539$\pm$0.033 &  $-$0.512 &    342 \\
 	  &   $-$9.72 & $-$0.063$\pm$0.041 &  $-$0.002 &    153 \\
          &	    &  		          &	      &        \\ 
\nuvr 	  &    1.62 &     0.174$\pm$0.050 & \phm 0.130 &     39 \\
 	  &    2.25 &  $-$0.090$\pm$0.028 &  $-$0.065 &    190 \\
 	  &    2.98 &  $-$0.593$\pm$0.030 &  $-$0.577 &    198 \\
 	  &    3.79 &  $-$0.987$\pm$0.033 &  $-$1.023 &    180 \\
 	  &    4.59 &  $-$1.281$\pm$0.040 &  $-$1.362 &    155 \\
  	  &    5.44 &  $-$1.514$\pm$0.026 &  $-$1.631 &    245 \\
  	  &    6.04 &  $-$1.672$\pm$0.020 &  $-$1.725 &    144 \\
\hline
\end{tabular}
\begin{flushleft}
Notes. $-$-- $^{a}$Weighted average of logarithm of gas fraction; \hi\ mass of non-detections set to upper limit.
$^{b}$Weighted median of logarithm of gas fraction; \hi\ mass of non-detections set to upper limit.
$^{c}$Number of galaxies in the bin.
\end{flushleft}
\end{table}

\hi\ gas fractions keep increasing with decreasing stellar mass, with no sign of a plateau,
down to \Mst $= 10^9$~\Msun\ (median values of \Mhi/\Mst\ increase from 2\% to 81\% from the highest
to the lowest stellar mass bin). This is consistent with the relation for \hi-rich galaxies detected 
by ALFALFA, which shows a flattening only below \Mst \about $10^{8.5}$~\Msun\ \citep{huang12}.
The correlation between \hi\ gas fraction and stellar mass has the largest scatter ($\bar{\Delta}=0.96$ dex). 
This is not surprising, as we already showed in previous work that variations of atomic gas fraction
at fixed stellar mass strongly correlate with star formation activity \citep{toby1}.
By applying spectral stacking to a large stellar mass-selected sample with \hi\ data from ALFALFA,
\citet[][{\bc see their Fig. 5}]{toby1} demonstrated that this relation is the result of a more physical correlation between
\hi\ content and SFR, combined with the bimodality of galaxies --- dividing up their sample
in three \nuvr\ color bins, roughly corresponding to blue sequence, red sequence and green valley,
they obtained three parallel relations with significantly flatter slope.

Interestingly, the relation with stellar surface density has lower scatter ($\bar{\Delta}=0.73$ dex), 
indicating that the correlation between gas fraction and stellar content improves when we take into account 
the size of the stellar disk (even if estimated as a 50\% effective radius). As noted before, a distinct 
difference between the \Mst\ and \must\ relations is the distribution of the non-detections, which are 
spread across the stellar mass range but pile up in the bulge-dominated region 
($\log$ \must [\Msun~kpc$^{-2}$] $\gtrsim 8.5$) -- above this threshold, both \hi\ and \htwo\ detection rates 
drop significantly \citep{gass1,coldgass1}. 
Galaxies in the lowest stellar surface density bin have median gas fractions of 100\%, \ie\ have the same amount
of mass in \hi\ gas and in stars; those with the bluest \nuvr\ colors are gas-dominated, reaching 
median gas fractions of 135\%.

\begin{figure}
\begin{center}
\includegraphics[width=7.5cm]{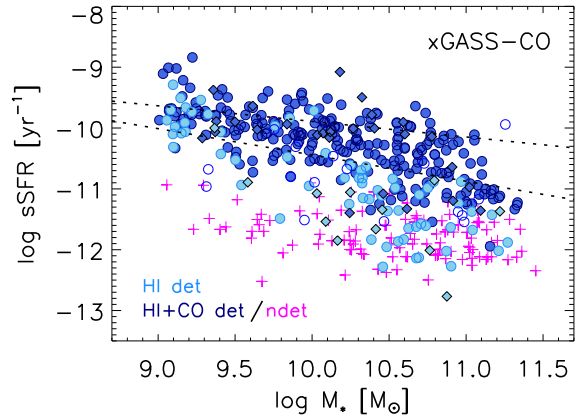}
\caption{Specific SFR plotted as a function of stellar mass for the subset of xGASS with CO data. Filled circles indicate galaxies
detected in both \hi\ and CO (dark blue) or \hi\ only (light blue); black-edged diamonds are \hi\ detections affected by beam
confusion. Empty circles are CO detections with \hi\ upper limits, and magenta crosses are non-detections in both 
\hi\ and CO lines. Dotted lines show the star-forming sequence adopted in this work (see text), and correspond to the 1$\sigma$ 
deviation above and below the average.
}
\label{xgass_co}
\end{center}
\end{figure}

\begin{figure*}
\begin{center}
\includegraphics[width=14cm]{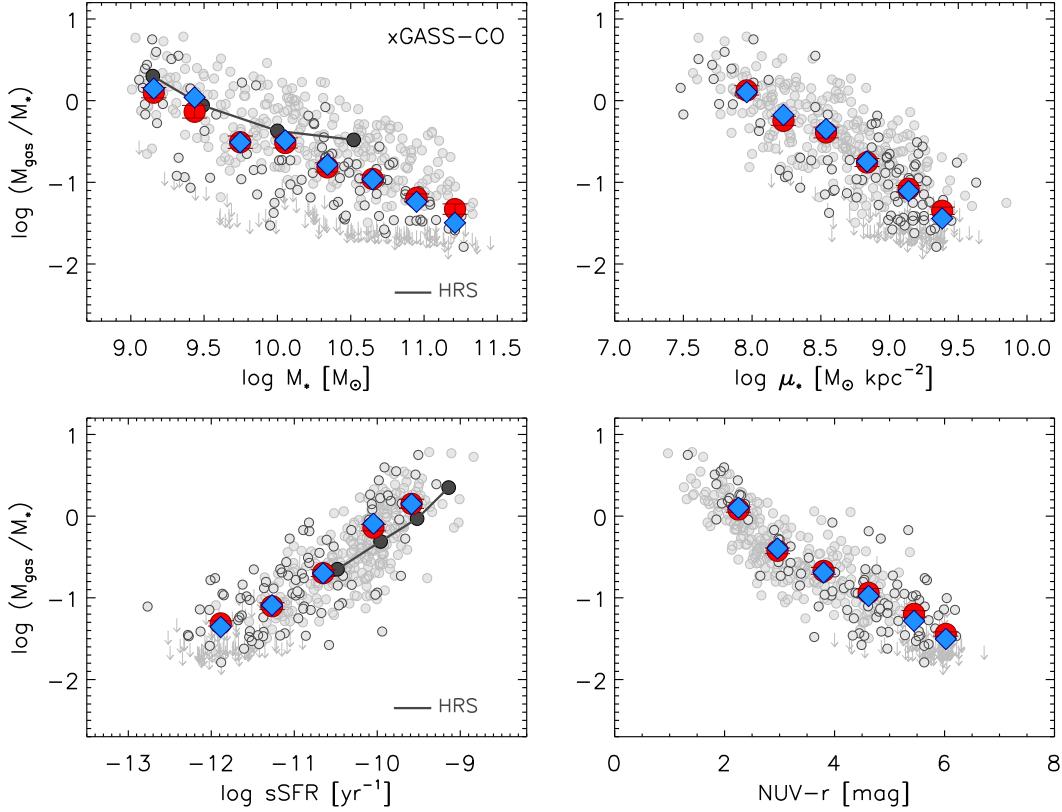}
\caption{Total gas fraction scaling relations for the xGASS-CO sample. Small gray circles and downward arrows show individual 
detections and non-detections in both \hi\ and CO lines; gray circles with darker contours indicate galaxies 
detected in \hi\ or CO, but not both. As in Figure~\ref{scalings_avgs}, large red circles and blue diamonds are weighted averages 
and weighted medians of the logarithms of the (total) gas fractions. These results are listed in Table~\ref{t_totgas_avgs}.
Black circles connected by lines in the left plots
show the results from the Herschel Reference Survey \citep[HRS;][]{boselli14b}.
}
\label{totalgas}
\end{center}
\end{figure*}

\subsection{Total gas fraction scaling relations}

In the rest of this paper we restrict our analysis to the subset of xGASS with IRAM observations, xGASS-CO, 
which includes 477 galaxies (see Section~\ref{s_xcoldgass}).

Figure~\ref{xgass_co} shows the distribution of this sample in the specific SFR versus stellar mass plane,
with points color-coded according to their detection status in the \hi\ and \htwo\ surveys. About 57\% of
the sample is detected in both lines (filled, dark blue circles), 16\% is detected only in \hi\ (light blue) 
and 23\% has no cold gas detection (crosses). As expected, galaxies on or above the star-forming main sequence 
(SFMS, dotted lines; see below) are typically \hi\ and \htwo\ detections, except at the low stellar mass end, where 
metallicities are lower and CO emission is more challenging to detect. We marked with black-edged diamonds
39 galaxies (8\% of xGASS-CO) with \hi\ emission that is confused within the Arecibo beam (see Appendices);
together with the non-detections in both gas phases, these objects are excluded from the analysis of 
{\bc molecular-to-atomic gas mass} ratios in the next section.
It is interesting to note that 17 galaxies (4\% of our sample, empty circles) are detected in CO only; 
of these, 9 are satellites in groups with 19 or more members according to the \citet{yang07} group catalog\footnote{
We use their SDSS DR7 ``B'' catalog, available online at {\it http://gax.shao.ac.cn/data/Group.html}; see
\citet{steven1} for more details.
}
and are all located in the bottom half of the SFMS or below it, suggesting that environmental effects in these large
groups might have depleted the \hi\ reservoirs, but not as far inside as the optical disk, thus leaving the \htwo\ content 
mostly unaffected \citep{fumagalli09,boselli14c,luca16a}.

We defined our own SFMS using the full xGASS representative sample. Briefly, we binned the points in the sSFR-
stellar mass plot (not shown) in stellar mass intervals of 0.25 dex, and fit Gaussians to the resulting sSFR 
distributions; this works well below \Mst \about $10^{10}$~\Msun, where the red sequence is almost absent. At 
higher stellar masses, we fit Gaussians with fixed centers based on the extrapolation of the relation at lower 
\Mst, and use only sSFRs above the relation to constrain the widths of the Gaussians. This procedure (illustrated 
in more detail in Janowiecki et al. in preparation) yields the following expression:
\begin{equation}
\log sSFR_{MS} =-0.344 (\log M_\star -9) -9.822
\label{eq_SFMS}
\end{equation}

\noindent
with a standard deviation given by:
\begin{displaymath}
\sigma_{MS}=0.088 (\log M_\star -9)+0.188
\end{displaymath}

\noindent
The limits corresponding to $\pm 1\sigma_{MS}$ from the SFMS are shown as dotted lines in Figure~\ref{xgass_co}.

Figure~\ref{totalgas} shows the scaling relations for the total gas (where $M_{\rm gas}=1.3(M_{\rm HI}+M_{\rm H2})$, 
including the Helium contribution). As in Figure~\ref{scalings_avgs}, weighted average (large red circles) and 
median (blue diamonds) gas fractions are plotted on top of individual measurements (small gray symbols; circles with 
darker contours are galaxies detected in either \hi\ or CO), with the same axis scales for comparison.
Median total gas fractions (computed including all detections and upper limits) decrease from 141\% to 3\%
over our stellar mass range; the galaxies with the highest total gas fractions in our sample have six times
more mass in cold gas than stars.

The observed trends are qualitatively similar to the ones seen for the atomic phase, but with slightly smaller scatter,
especially in the relations between gas fraction and stellar mass or stellar surface density. If we quantify the
dispersions of these relations with the parameter $\bar{\Delta}$ defined in the previous section, \ie\ the average
difference between the 75th and 25th percentiles of the $\log$(\Mgas/\Mst) distributions in each bin, we
obtain $\bar{\Delta}=0.70, 0.61, 0.47$ and 0.36 dex for the \Mst, \must, sSFR and \nuvr\ relations.
These should be compared with the dispersions of the \hi\ scaling relations computed for the same xGASS-CO sample,
which are $\bar{\Delta}=0.81, 0.69, 0.50$ and 0.40 dex, respectively. This difference is most likely due to the
fact that the total gas fractions have smaller dynamic range than the atomic ones.

Overall, the similarity between atomic and total gas scaling relations is not surprising,
as galaxies in this stellar mass regime in the local Universe typically have cold gas reservoirs that are \hi-dominated
\citep[][see also next section]{coldgass1,boselli14b,saintonge16}.

Our total gas scaling relations confirm and extend to higher stellar mass the results of \citet{boselli14b}, 
obtained for field late-type galaxies detected in both \hi\ and CO lines in the Herschel Reference Survey \citep[HRS;][]{hrs},
assuming a luminosity-dependent $\rm X_{CO}$ conversion factor to compute molecular gas masses. {\bc For reference, the 
HRS results are shown as black circles connected by lines in Fig.~\ref{totalgas} (left panels).}
The agreement with their stellar mass relation in the overlap interval ($\log$\Mst~[\Msun] $< 10.5$) is excellent,
except for their highest stellar mass bin, which has a total gas fraction 0.4 dex higher than ours, probably due
to limited statistics of the HRS at the high \Mst\ end. The relation with sSFR for the HRS galaxies has the same slope
but is slightly offset towards lower gas fractions (by \about 0.2 dex); however the two samples overlap by only 
\about 1.5 dex in sSFR.

Interestingly, \citet{boselli14b} noted that the HRS relationships involving molecular gas fractions are always flatter  
than those with total gas fraction. This is confirmed by our sample \citep{accurso17,xcoldgass}.
Indeed, thanks to the larger dynamic range in \Mst\ and \must\ of xCOLD GASS, we detect a clear break in
the molecular gas relations, which suddenly flatten below $\log$~\Mst [\Msun] $= 10.5$ and 
$\log$~\must [\Msun~kpc$^{-2}$] $= 8.5$ \citep[see][]{accurso17,xcoldgass}. As seen in Figure~\ref{totalgas}, 
there is no trace left of such flattening in the total gas relations.
The difference between atomic and molecular gas fraction relations below these stellar mass and stellar
surface density limits is striking, and warrants a closer look at the molecular-to-atomic mass ratio in the next section.

\begin{figure}
\begin{center}
\includegraphics[width=7.3cm]{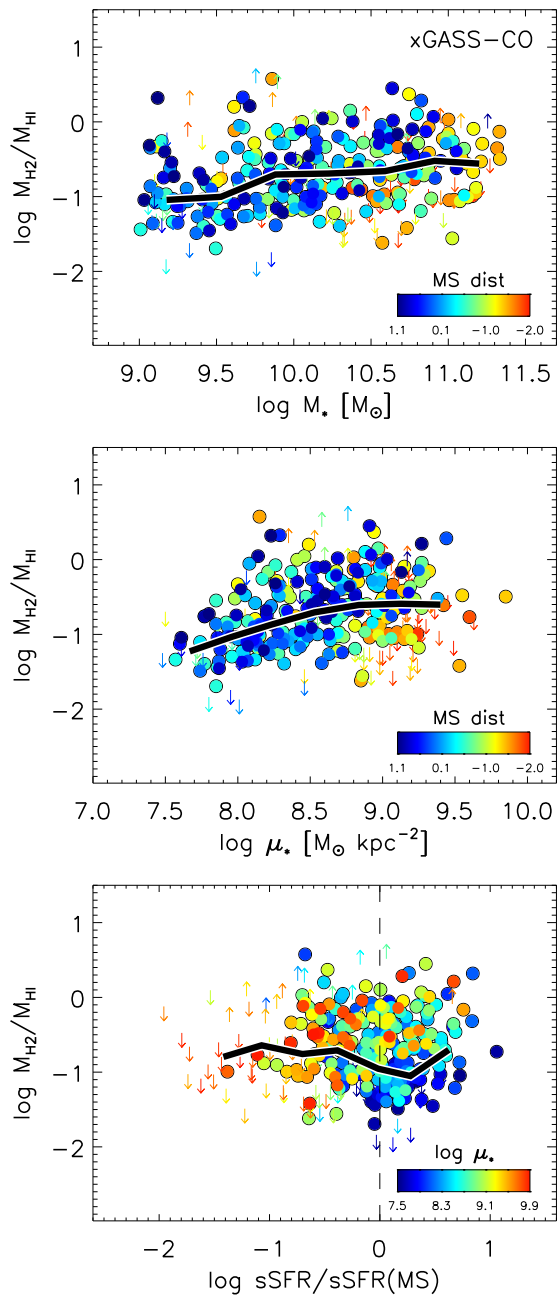}
\caption{Molecular-to-atomic gas mass ratio as a function of stellar mass (top), stellar surface density (middle)
and deviation from the star-forming main sequence (bottom). In each panel, circles are galaxies with both \hi\ and CO
detections, downward arrows are \hi\ detections with CO upper limits, and upward arrows are CO detections with \hi\
upper limits. Galaxies are color-coded {\bc by} distance from the SF sequence in the top two panels, and
{\bc by} \must\ in the bottom one. Thick lines show running medians; only medians computed with at least 10 galaxies are shown.
}
\label{rmol}
\end{center}
\end{figure}

\subsection{Molecular-to-atomic gas mass ratios}

In our previous work we investigated the relation between \hi\ and \htwo\ content for the initial release of the 
GASS+COLD GASS sample, and found that the {\bc molecular-to-atomic gas mass} ratio, $R_{\rm mol} \equiv M_{\rm H2}/M_{\rm HI}$, weakly increases with
stellar mass, stellar surface density and \nuvr\ color, but with over 0.4 dex of scatter \citep{coldgass1}. 
We also showed how \rmol\ varies across the SFR-\Mst\ plane for the full GASS+COLD GASS sample,
and identified a region of unusually high values of {\bc \rmol\ ($>0.7$)} at high stellar masses and
SFRs ($\log M_{\star} [M_{\odot}]> 10.8$ and $\rm -10.4 < \log~sSFR [yr^{-1}] <-9.6$; \citealt{saintonge16}). These galaxies 
are characterized by young stellar populations in their central regions (based on their $\rm D_n4000$ from SDSS fiber 
spectroscopy) and important bulge components ($\log \mu_{\star} \sim 8.9$).
Here we extend these studies to lower stellar mass, focusing on the relation between galaxies within or outside the
SFMS defined in the previous section. Galaxies with non-detections in both \hi\ and \htwo\ (magenta crosses in Fig.~\ref{xgass_co})
are excluded from this analysis, because their \rmol\ is unconstrained; we also exclude \hi-confused detections
(black-edged diamonds in Fig.~\ref{xgass_co}), which leaves us with a sample of 328 galaxies.

\begin{figure*}
\begin{center}
\includegraphics[width=14cm]{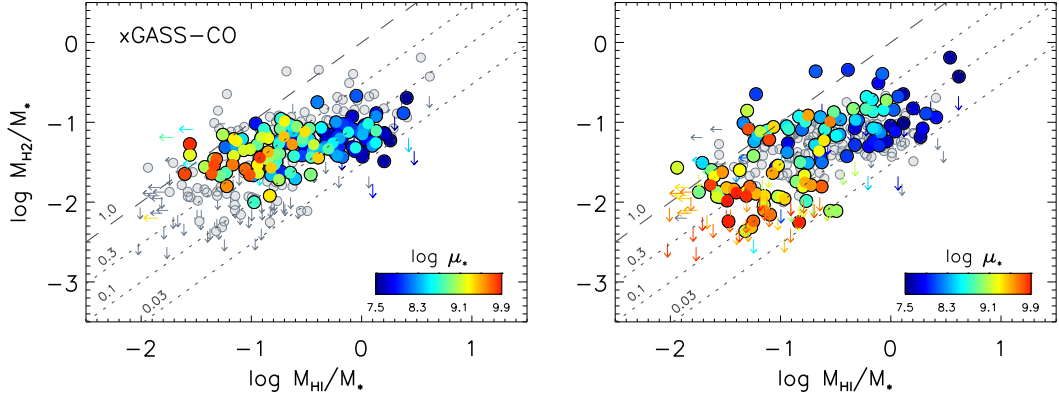}
\caption{Comparison between \htwo\ and \hi\ gas fractions. Circles are galaxies detected in both
\hi\ and \htwo; downward and leftward arrows are \htwo\ and \hi\ upper limits, respectively. 
The two panels highlight galaxies within 1$\sigma$ from the SFMS (left; see Fig.~\ref{xgass_co}) and 
outside {\bc the SFMS} (right panel), color-coded according to their stellar surface density as indicated.
{\bc Gray symbols in each panel show the excluded galaxies (\ie, correspond to the colored points in the other panel).} 
Also shown are lines of constant {\bc molecular-to-atomic gas mass} ratio, with \Mhtwo/\Mhi\ values as labeled.
}
\label{gasfr}
\end{center}
\end{figure*}

Figure~\ref{rmol} (top two panels) shows the {\bc molecular-to-atomic gas mass} ratio as a function of stellar mass and stellar surface density 
for our sample; thick lines indicate running weighted medians of the logarithm of \rmol. The median \rmol\ 
slowly increases with both stellar mass (from 9\% to 28\%) and stellar surface density (from 6\% to 25\%; see Table~\ref{t_rmol_avgs}). 
A similar trend with stellar mass was also found by the APEX Low-redshift Legacy Survey for MOlecular Gas 
\citep[ALLSMOG,][]{allsmog} and the HRS \citep{boselli14b}, which both combined COLD GASS data with new observations
probing stellar masses below $10^{10} M_{\odot}$.
Furthermore, galaxies in the top two panels of Figure~\ref{rmol} are color-coded according to their distance from the SFMS 
(\ie, $\rm dist_{MS}=\log sSFR - \log sSFR_{MS}$, see equation~\ref{eq_SFMS}); negative values (redder colors) correspond to systems below the SFMS.
Galaxies located below the SFMS typically have high 
stellar surface densities; this is better seen in the bottom panel, where \rmol\ is plotted as a function of distance from the SFMS, 
and color-coded by \must. On and above the SFMS (dashed line), bulge-dominated systems are displaced towards higher {\bc \rmol\ values}.

The scatter in these plots is quite large, with \Mhtwo/\Mhi\ ratios that vary by almost two orders of magnitude across our sample.
As seen in the middle panel, galaxies below the SFMS (which are typically bulge-dominated) seem to follow a different relation from
star-forming disks. This qualitatively agrees with the observation that HRS early-type galaxies detected in \hi\ do not
follow the same gas scaling relations as late-type ones \citep{boselli14b}. However, it is unclear if the observed scatter
correlates more strongly with deviation from the SFMS or stellar surface density. 

In order to gain further insight into what regulates the molecular-to-atomic {\bc gas} mass ratio of our sample, we compare
\hi\ and \htwo\ gas fractions directly in Figure~\ref{gasfr}. In the left panel, galaxies on the SFMS are color-coded
by stellar surface density; the right panel shows the complementary set of galaxies located outside the SFMS, with the same
color coding. Looking at the right panel first, there is a general trend of increasing molecular gas fractions for 
increasing \Mhi/\Mst, with a clear dependence on stellar surface density. As can be seen, more bulge-dominated 
systems (redder colors in the figure) have systematically lower atomic and molecular gas fractions, while spanning the 
full range of \rmol. Very interestingly, and contrary to the rest of the sample, the relation for SFMS galaxies (left) is nearly 
flat -- selecting galaxies within 1$\sigma$ of the SFMS restricts \Mhtwo/\Mst\ to vary within a dex, whereas atomic gas fractions 
still span almost the entire range of the full sample. This is highlighted by the lines of constant {\bc molecular-to-atomic gas mass} 
ratio, which increases from 3\% to 100\% from bottom to top, and shows that {\it the observed variation of \rmol\ is mostly driven by 
changes of the atomic gas reservoir} -- not the molecular one. 

\begin{figure*}
\begin{center}
\includegraphics[width=14cm]{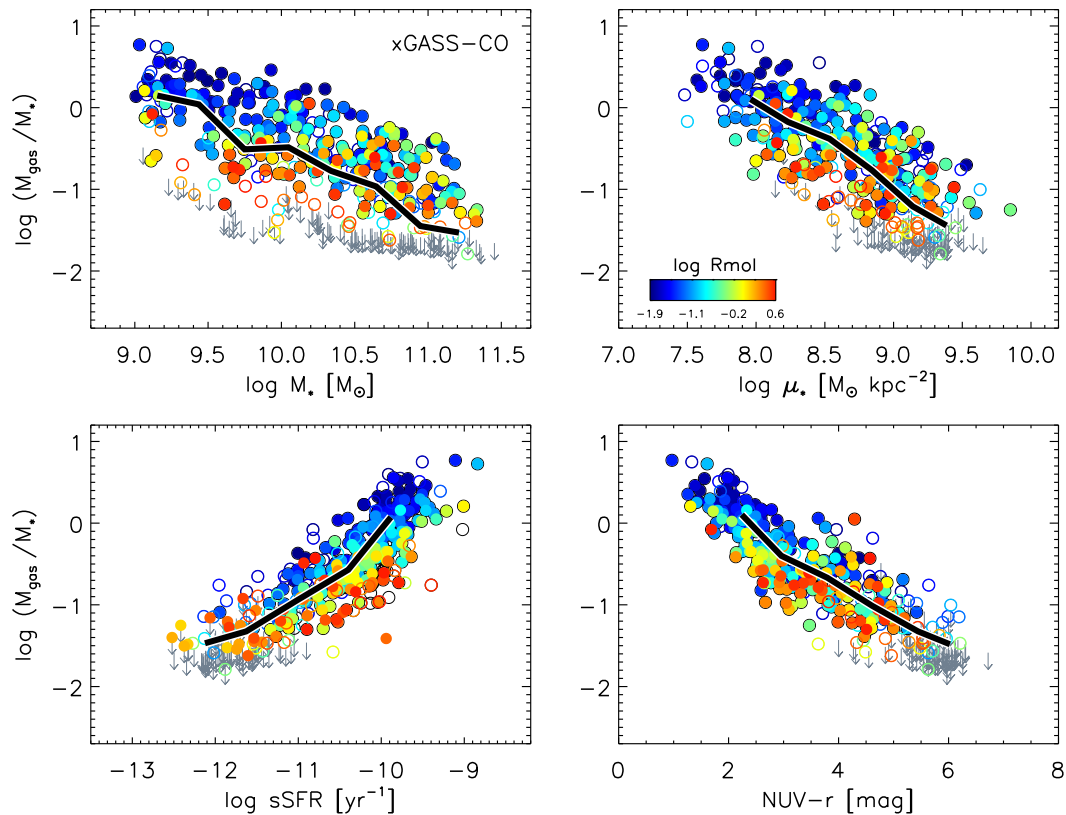}
\caption{Total gas fraction scaling relations, color-coded by {\bc molecular-to-atomic gas mass} ratio. Empty circles indicate galaxies not detected in either 
\hi\ or \htwo, and gray downward arrows are systems with upper limits in both gas phases, for which \rmol\ is not defined. 
This sample does not include \hi-confused galaxies, hence running medians (thick lines) are not identical to those 
in Figure~\ref{totalgas} (blue diamonds).
}
\label{totalgas_rmol}
\end{center}
\end{figure*}

This finding suggests that the {\bc scatter in} the total gas scaling relations might also be related to the variation of
molecular-to-atomic {\bc gas} mass ratio, which is indeed the case. This is demonstrated in Figure~\ref{totalgas_rmol}, which presents
the relations of Figure~\ref{totalgas} with points color-coded by \rmol; gray arrows are galaxies with upper limits
in both \hi\ and \htwo\, for which \rmol\ is not defined. The variation of total gas fraction is clearly driven by 
a change in {\bc molecular-to-atomic gas mass} ratio in all these plots. The secondary dependence on \rmol\ is most prominent 
{\it at fixed specific SFR}, where galaxies with smaller total gas reservoirs have larger {\bc values of \rmol}.

Lastly, the left panel of Figure~\ref{rmol} shows that, because \htwo\ gas fractions are to first order roughly constant on the SFMS, 
the decrease of \hi\ gas fractions leads to higher \rmol\ for bulge-dominated systems, as observed in Figure~\ref{rmol}.
It is tempting to interpret these trends with stellar stellar surface density as suggestive of a causal link between galaxy structure 
and gas content. However, we obtain very similar results if we color-code the galaxies in this figure by stellar mass (not shown), 
pointing out the difficulty of separating the effects of mass and structure using global measurements.

\begin{table}
\centering
\caption{Total gas fraction scaling relations for xGASS-CO}
\label{t_totgas_avgs}
\begin{tabular}{lrrcr}
\hline\hline
$x$  & $\langle x \rangle$ & $\langle M_{\rm gas}/M_\star \rangle^{a}$  & $(M_{\rm gas}/M_\star)^{b}$   &  $N^{c}$ \\
\hline
log \Mst  &    9.16 &  \phm 0.098$\pm$0.064 & \phm 0.148 &     41 \\
 	  &    9.44 &  $-$0.136$\pm$0.077 & \phm 0.040 &     43 \\
 	  &    9.75 &  $-$0.509$\pm$0.076 &   $-$0.511 &     54 \\
 	  &   10.05 &  $-$0.518$\pm$0.062 &   $-$0.485 &     69 \\
 	  &   10.34 &  $-$0.817$\pm$0.055 &   $-$0.785 &     75 \\
 	  &   10.65 &  $-$0.958$\pm$0.048 &   $-$0.965 &     89 \\
 	  &   10.95 &  $-$1.190$\pm$0.048 &   $-$1.238 &     74 \\
 	  &   11.21 &  $-$1.328$\pm$0.064 &   $-$1.496 &     31 \\
          &	    &  		          &	       &        \\ 
log \must &    7.96 &  \phm 0.125$\pm$0.057 & \phm 0.104 &     44 \\
 	  &    8.23 &  $-$0.247$\pm$0.056 &   $-$0.176 &     65 \\
 	  &    8.54 &  $-$0.390$\pm$0.067 &   $-$0.348 &     64 \\
 	  &    8.84 &  $-$0.757$\pm$0.049 &   $-$0.746 &    110 \\
 	  &    9.14 &  $-$1.079$\pm$0.040 &   $-$1.106 &    130 \\
 	  &    9.38 &  $-$1.348$\pm$0.045 &   $-$1.443 &     46 \\
          &	    &  		          &	       &        \\ 
log sSFR  & $-$11.89 &  $-$1.313$\pm$0.032 &   $-$1.356 &     87 \\
 	  & $-$11.27 &  $-$1.102$\pm$0.042 &   $-$1.089 &     91 \\
 	  & $-$10.65 &  $-$0.698$\pm$0.042 &   $-$0.699 &     88 \\
 	  & $-$10.05 &  $-$0.141$\pm$0.039 &   $-$0.092 &    149 \\
 	  &  $-$9.59 &  \phm 0.151$\pm$0.056 & \phm 0.150 &     44 \\
          &	    &  		          &	       &        \\  
\nuvr     &    2.25 &     0.082$\pm$0.034 & \phm 0.104 &     95 \\
 	  &    2.96 &  $-$0.424$\pm$0.035 &   $-$0.394 &     80 \\
 	  &    3.80 &  $-$0.670$\pm$0.042 &   $-$0.689 &     69 \\
 	  &    4.62 &  $-$0.938$\pm$0.052 &   $-$0.978 &     76 \\
 	  &    5.44 &  $-$1.197$\pm$0.042 &   $-$1.282 &     82 \\
 	  &    6.02 &  $-$1.442$\pm$0.026 &   $-$1.504 &     52 \\
\hline
\end{tabular}
\begin{flushleft}
Notes. $-$-- $^{a}$Weighted average of logarithm of gas fraction; \hi\ mass of non-detections set to upper limit.
$^{b}$Weighted median of logarithm of gas fraction; \hi\ mass of non-detections set to upper limit.
$^{c}$Number of galaxies in the bin.
\end{flushleft}
\end{table}

\begin{table}
\centering
\caption{Molecular-to-atomic gas mass ratio scaling relations for xGASS-CO}
\label{t_rmol_avgs}
\begin{tabular}{lrrcr}
\hline\hline
$x$  & $\langle x \rangle$ & $\langle$\Mhtwo/\Mhi $\rangle^{a}$  & (\Mhtwo/\Mhi)$^{b}$   &  $N^{c}$ \\
\hline
log \Mst  &    9.18 &  $-$0.903$\pm$0.070 &   $-$1.044 &     44 \\
 	  &    9.53 &  $-$0.865$\pm$0.079 &   $-$1.003 &     38 \\
 	  &    9.88 &  $-$0.660$\pm$0.073 &   $-$0.704 &     53 \\
 	  &   10.22 &  $-$0.704$\pm$0.059 &   $-$0.691 &     59 \\
 	  &   10.59 &  $-$0.662$\pm$0.057 &   $-$0.659 &     65 \\
 	  &   10.91 &  $-$0.530$\pm$0.058 &   $-$0.521 &     50 \\
 	  &   11.20 &  $-$0.578$\pm$0.075 &   $-$0.559 &     18 \\
          &         &		          & 	       &        \\ 
log \must &    7.67 &  $-$1.132$\pm$0.066 &   $-$1.228 &     14 \\
 	  &    7.96 &  $-$1.018$\pm$0.069 &   $-$1.032 &     41 \\
 	  &    8.23 &  $-$0.682$\pm$0.066 &   $-$0.870 &     57 \\
 	  &    8.53 &  $-$0.683$\pm$0.072 &   $-$0.706 &     49 \\
 	  &    8.84 &  $-$0.583$\pm$0.060 &   $-$0.606 &     79 \\
 	  &    9.13 &  $-$0.603$\pm$0.054 &   $-$0.591 &     70 \\
 	  &    9.40 &  $-$0.602$\pm$0.122 &   $-$0.603 &     14 \\
\hline
\end{tabular}
\begin{flushleft}
Notes. $-$-- $^{a}$Weighted average of logarithm of \rmol; \hi\ and \htwo\ masses of non-detections set to upper limits.
$^{b}$Weighted median of logarithm of \rmol; \hi\ and \htwo\ masses of non-detections set to upper limits.
$^{c}$Number of galaxies in the bin.
\end{flushleft}
\end{table}


\section{Comparison with models}\label{s_models}

Gas fraction scaling relations for representative samples have a unique constraining power for galaxy
formation models \citep[\eg,][]{lagos14,lagos15,popping14,popping15,bahe16,mufasa_gas}. 
Modern cosmological semi-analytic and hydrodynamical simulations successfully reproduce overall
stellar and star formation properties of galaxies over cosmic time, but must rely on sub-resolution
prescriptions to partition the cold gas into atomic and molecular phases and form stars. Comparisons between observed gas 
scaling relations from stellar mass-selected samples and simulated ones have highlighted areas where models need
improvement \citep[\eg,][]{gk12,toby2,stevens17,zoldan17}. 

In our companion paper \citep{xcoldgass}, we compared \hi\ and \htwo\ gas fractions as a function of stellar mass
with predictions from two large, state-of-the-art hydrodynamical simulations,
MUFASA \citep{mufasa,mufasa_gas} and the Evolution and Assembly of GaLaxies and their Environments 
(EAGLE, \citealt{eagle}. We used their high resolution {\it Recal-L025N0752} run). 

Briefly, MUFASA directly tracks the amount of molecular gas formed in galaxies
using a sub-resolution prescription (broadly following \citealt{krumholz09}). The atomic fraction is obtained
by subtracting the molecular fraction from the neutral (self-shielded against the cosmic metagalactic flux) gas,
and the global \hi\ content of a galaxy is just the sum of the atomic gas that is bound to it.

For the EAGLE simulations, the partition of the ISM into its different phases was implemented by \citet{lagos15} 
in post-processing. The separation between ionized and neutral (self-shielded) gas is done according to the same prescription
adopted by MUFASA (based on \citealt{rahmati13}); the neutral gas is then divided into \hi\ and \htwo\ phases
following \citet[][GK11]{gnedin11} or \citet[][K13]{krumholz13}. Both recipes give \htwo\ fractions that depend 
on gas metallicity and strength of the interstellar radiation field, but the partition into \hi\ and \htwo\  
phases relies on the assumption that the warm and cold components of the ISM are in pressure equilibrium (K13) 
or is based on metallicity, since \htwo\ formation happens on dust grains (GK11). 

The comparison with our results showed that both MUFASA and EAGLE simulations reproduce reasonably well the 
\hi\ gas fractions in galaxies with $\log M_{\star} [M_{\odot}] < 10.5$, but significantly underpredict the 
amount of cold atomic gas in more massive galaxies. Contrary to the \hi\ phase, predictions of \htwo\ gas
fractions are very sensitive to the subgrid physics assumed to partition the ISM, and we found that none
of these hydrodynamical simulations reproduce the molecular gas content of galaxies with 
$\log M_{\star} [M_{\odot}] < 10.5$ particularly well \citep{xcoldgass}. The best agreement is with the 
EAGLE K13 prescription, whereas EAGLE GK11 and MUFASA produce galaxies with too much molecular gas.

\begin{figure}
\begin{center}
\includegraphics[width=7.5cm]{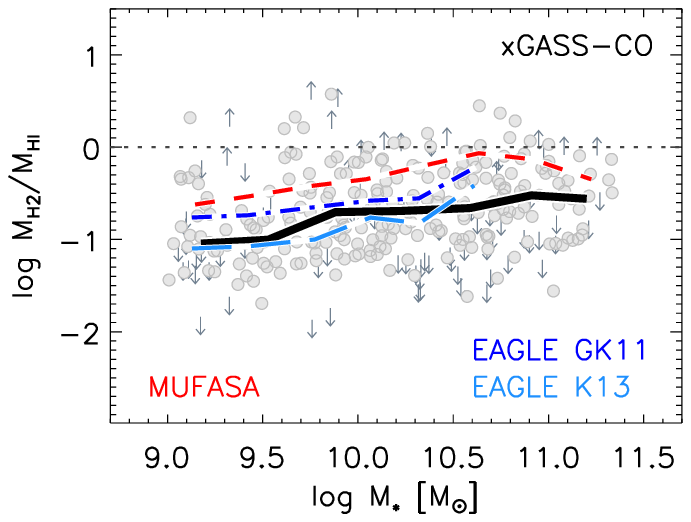}
\caption{Comparison with hydrodynamical simulations. The molecular-to-atomic gas mass ratio as a 
function of stellar mass is reproduced from the top panel of Fig.~\ref{rmol} (gray points), along
with the median relation for our sample (black line). Predictions from MUFASA \citep{mufasa} and EAGLE \citep{lagos15}
hydrodynamical models are shown as red and blue lines, respectively. The two EAGLE models differ for the 
prescription adopted to partition the cold gas into atomic and molecular phases (GK11: \citealt{gnedin11}; 
K13: \citealt{krumholz13}). The horizontal dotted line is at \Mhtwo/\Mhi\ $=1$, separating atomic- from 
molecular-dominated systems.
}
\label{rmol_sims}
\end{center}
\end{figure}

Interestingly, despite the fact that \hi\ and \htwo\ gas fractions are not individually well reproduced by 
{\bc these hydrodynamical simulations} across the full stellar mass range of our sample, Figure~\ref{rmol_sims} shows that 
{\bc the molecular-to-atomic gas mass {\it ratio} predicted by EAGLE}
is in overall better agreement with our observations (gray symbols, with the solid black 
line showing the median relation). In this figure, red and blue lines indicate median {\bc \rmol\ values} from MUFASA and 
{\it Recal-L025N0752} EAGLE run respectively; for the latter, light and dark blue lines correspond
to the K13 and GK11 prescriptions. In order to be consistent with our observations, we applied our gas fraction
limits to the simulated data sets, and excluded galaxies that would not be detected in both \hi\ and \htwo\
before computing the medians. MUFASA galaxies have {\bc molecular-to-atomic gas mass} ratios \about 0.4 dex higher than observed.
{\bc This is because, to partly compensate for its lower resolution, MUFASA effectively employs a lowered 
density threshold for forming \htwo, which results in more ISM gas being molecular rather than 
atomic at a given stellar mass.}
The EAGLE K13 model provides an excellent match to our data, whereas the GK11
version slightly {\bc but} systematically overestimates \rmol. We note that, while
\hi\ gas fractions above $\log M_{\star} [M_{\odot}]$\about 10.2, are similarly underestimated in both subgrid 
implementations, the \htwo\ fractions are underestimated by K13 (thus getting \rmol\ approximately correct) and 
overestimated for GK11, resulting in higher {\bc molecular-to-atomic gas mass} ratios. This shows the importance of testing multiple
gas scaling relations to constrain the physics implemented in simulations.

\section{Summary and conclusions}\label{s_concl}

In this paper we presented xGASS, the culmination of several years of effort to gather deep \hi\ observations
for a stellar mass-selected sample of \about 1200 galaxies with homogeneously measured optical and SF properties.
xGASS is the combination of the original GASS survey, which started in 2008 and targeted galaxies with stellar 
masses larger than $10^{10}$~\Msun, and its extension to $M_\star = 10^9$~\Msun. Together, these surveys required
\about 1300 hours of Arecibo telescope time.
Our unique approach of carrying out gas fraction-limited observations down to \Mhi/\Mst \about 2\% allowed us to
obtain stringent upper limits, which are essential to interpret variations of gas content as a function of
galaxy properties. 

We release here \hi\ catalogs and spectra for the complete low mass extension of GASS, which includes new Arecibo 
observations of 208 galaxies. By adding the correct proportion of ALFALFA \hi-rich systems (not targeted by us to 
increase survey efficiency) to GASS and GASS-low data sets, we obtained a representative sample (in terms of \hi\ content)
of {\bc 1179} galaxies with stellar mass $10^9 \leq M_\star/ M_\odot \leq 10^{11.5}$ in the local Universe ($0.01 \leq z \leq 0.05$).

In addition to extending the \hi\ scaling relations by one decade in stellar mass, we quantified total gas fraction 
scaling relations for the subset of 477 galaxies {\bc with molecular hydrogen mass estimates available}, and explored
molecular-to-atomic gas mass variations for galaxies detected in at least one of the two gas phases. 
\htwo\ masses were obtained as part of the xCOLD GASS follow-up survey, which measured the CO(1-0) line emission 
of xGASS galaxies using the IRAM 30m radio telescope \citep{xcoldgass}. Our main results are summarized below.

\begin{itemize}

\item Atomic gas fractions increase from 2\% (set by the limit of our observations) to 81\% with decreasing stellar mass,                      
with no sign of a plateau. The tightest relation is with \nuvr\ color, which traces dust-unobscured star formation (as opposed
to molecular gas fractions that correlate more strongly with sSFR, \citealt{xcoldgass}).

\item On average, galaxies have gas reservoirs that remain dominated by atomic hydrogen across the full range of stellar 
masses probed by our survey (see also Fig.~\ref{rmol_sims}). Molecular-to-atomic hydrogen mass ratios weakly
increase with stellar mass from 9\% to 27\%, but varying by two orders of magnitude across the sample.

\item Total gas fraction scaling relations closely resemble atomic ones, as expected from the fact that \hi\ is the
dominant gas phase. Below $\log M_\star/ M_\odot \sim 9.5$, the median galaxy has more mass in cold gas than stars.
The {\bc scatter in} the total gas fraction relations is driven by changes in \rmol. At
fixed specific SFR, galaxies with larger total gas reservoirs have smaller {\bc molecular-to-atomic gas mass} ratios.

\item  For galaxies on the star-forming sequence, variations of \rmol\ are mostly driven by changes 
of the \hi\ reservoirs, with a clear dependence on stellar surface density. Bulge-dominated systems have \Mhtwo/\Mhi\ ratios
that are typically three times larger than those of disk-dominated galaxies. This highlights once again the importance 
of galaxy structure, as traced by stellar surface density, in relation 
to the cold gas content of galaxies \citep[see also][]{gass1,coldgass1,saintonge12,toby1}. 

\end{itemize}

When interpreting these results, one has to bear in mind that \hi\ and \htwo\ line fluxes are measured with radio
telescopes with vastly different beams (\about 3.5 arcmin for Arecibo and \about 22 arcsec for IRAM). We apply aperture
corrections to recover global \htwo\ masses, but nonetheless it is well known that most of the \hi\ is distributed
in the outer parts of galaxy disks, beyond the \htwo -dominated regions. Thus our \Mhtwo/\Mhi\ ratios carry information
on the {\it global} \hi\ and \htwo\ gas reservoirs available for future star formation, more than on the detailed conversion 
between the two. 
Even with this caveat, it remains very intriguing to investigate the reason(s) for the systematic variation of the 
molecular-to-atomic {\bc gas mass} ratio with stellar surface density, and ultimately presence of a bulge component. While it remains
difficult to establish if stellar mass or structure is more important in connection with the gas content of galaxies 
(at least using global quantities), there is no doubt that part of the scatter in all the relations presented in this work 
must be due to the fact that we normalize gas masses by stellar mass, which includes the bulge component, 
whereas the gas is found in the disk. We will address this issue in future work, by performing accurate {\bc photometric 
bulge-to-disk decompositions for xGASS galaxies to separate the total stellar mass into bulge and disk contributions,
$M_{\star,B}$ and $M_{\star,D}$.
We will then be able to investigate gas fraction scaling relations for the disk component alone (\ie, plotting $M_{\rm HI}/M_{\star,D}$)
and determine if and how these are affected by the presence of a bulge.}

Statistical measurements of the cold gas content for stellar mass-selected samples are a crucial test-bed for models of
galaxy formation. We presented an example by comparing {\bc molecular-to-atomic gas mass} ratios measured from our sample with two state-of-the-art
hydrodynamical simulations, MUFASA and EAGLE, and noted how sometimes good agreement is obtained overall, even though 
the underlying distributions are not well reproduced. This is a complex parameter space, with several systematic trends 
that are still not completely understood, thus it is essential to test simulations with the largest possible combination 
of ISM components and galaxy properties -- something that our large, homogeneous and very sensitive xGASS and xCOLD GASS 
surveys were precisely designed to provide.


\section*{Acknowledgments}

We thank Claudia Lagos for making the results of her simulations available and for useful discussions,
{\bc and an anonymous referee for a very careful reading of our paper and constructive comments.}
BC is the recipient of an Australian Research Council Future Fellowship (FT120100660). 
BC, SJ and LC acknowledge support from the Australian Research Council's Discovery
Projects funding scheme (DP150101734). APC acknowledges the support of STFC grant ST/P000541/1.

This research has made use of the NASA/IPAC Extragalactic Database
(NED) which is operated by the Jet Propulsion Laboratory, California
Institute of Technology, under contract with the National Aeronautics
and Space Administration.

The Arecibo Observatory is operated by SRI International under a
cooperative agreement with the National Science Foundation
(AST-1100968), and in alliance with Ana G. M{\'e}ndez-Universidad
Metropolitana, and the Universities Space Research Association.

GALEX (Galaxy Evolution Explorer) is a NASA Small Explorer, launched
in April 2003. We gratefully acknowledge NASA's support for
construction, operation, and science analysis for the GALEX mission,
developed in cooperation with the Centre National d'Etudes Spatiales
(CNES) of France and the Korean Ministry of Science and Technology. 

Funding for the SDSS and SDSS-II has been provided by the Alfred
P. Sloan Foundation, the Participating Institutions, the National
Science Foundation, the U.S. Department of Energy, the National
Aeronautics and Space Administration, the Japanese Monbukagakusho, the
Max Planck Society, and the Higher Education Funding Council for
England. The SDSS Web Site is http://www.sdss.org/.

The SDSS is managed by the Astrophysical Research Consortium for the
Participating Institutions. The Participating Institutions are the
American Museum of Natural History, Astrophysical Institute Potsdam,
University of Basel, University of Cambridge, Case Western Reserve
University, University of Chicago, Drexel University, Fermilab, the
Institute for Advanced Study, the Japan Participation Group, Johns
Hopkins University, the Joint Institute for Nuclear Astrophysics, the
Kavli Institute for Particle Astrophysics and Cosmology, the Korean
Scientist Group, the Chinese Academy of Sciences (LAMOST), Los Alamos
National Laboratory, the Max-Planck-Institute for Astronomy (MPIA),
the Max-Planck-Institute for Astrophysics (MPA), New Mexico State
University, Ohio State University, University of Pittsburgh,
University of Portsmouth, Princeton University, the United States
Naval Observatory, and the University of Washington.

\bibliography{biblio}


\section*{Appendix A: Data Release}

\setcounter{table}{0}
\renewcommand{\thetable}{A\arabic{table}} 

\setcounter{figure}{0}
\renewcommand{\thefigure}{A\arabic{figure}} 

We present here SDSS postage stamp images, Arecibo \hi-line spectra,
and catalogs of optical, UV and \hi\ parameters for the 208 GASS-low galaxies. 
The content of the tables is described below; notes on individual objects
(marked with an asterisk in the last column of Tables~\ref{t_det} and
\ref{t_ndet}) are reported in Appendix~B.\\

\noindent
{\bf SDSS and GALEX data.}\\
Table~\ref{t_sdss} lists optical and UV quantities for the 208 GASS-low
galaxies, ordered by increasing right ascension:\\

\bct
Cols. 1 and 2: GASS and SDSS identifiers. Galaxies with six digit GASS IDs are
part of GASS-low.\\

\bct
Col. 3: UGC \citep{ugc}, NGC \citep{ngc} or IC \citep{ic,ic2}
designation, or other name, typically from
the Catalog of Galaxies and Clusters of Galaxies \citep[CGCG;][]{cgcg}, 
or the Virgo Cluster Catalog \citep[VCC;][]{vcc}.\\

\bct
Col. 4: SDSS redshift, $z_{\rm SDSS}$. The typical uncertainty of
SDSS redshifts for this sample is 0.0002.\\

\bct
Col. 5: base-10 logarithm of the stellar mass, \Mst, in solar
units. Stellar masses are obtained from the SDSS DR7 MPA/JHU catalog
(see footnote 2 in section 2.3) and assume a \citet{chabrier03}
initial mass function. Over our stellar mass range, these values are
believed to be accurate to better than 30\%.\\

\bct
Col. 6: radius containing 50\% of the Petrosian flux in \zband, \Rinz, in arcsec.\\

\bct
Cols. 7 and 8: radii containing 50\% and 90\% of the Petrosian
flux in \rband, $R_{50}$ and  $R_{90}$ respectively, in arcsec.\\

\bct
Col. 9: base-10 logarithm of the stellar mass surface density, \must, in
\Msun~kpc$^{-2}$. This quantity is defined as $\mu_\star = M_\star/(2 \pi R_{50,z}^2)$, 
with \Rinz\ in kpc units (computed using angular distances).\\

\bct
Col. 10: Galactic extinction in \rband, ext$_r$, in magnitudes, from SDSS.\\

\bct
Col. 11: \rband\ model magnitude from SDSS, $r$, corrected for Galactic extinction.\\

\bct
Col. 12: minor-to-major axial ratio from the exponential fit in \rband, $(b/a)_r$, from SDSS.\\

\bct
Col. 13: inclination to the line-of-sight, in degrees (see \citealt{gass_dr2} for details).\\

\bct
Col. 14: \nuvr\ observed color, corrected for Galactic extinction, in magnitudes \citep[see][]{steven1}.\\

\bct
Col. 15: star formation rate, SFR, from NUV and WISE photometry, in \Msun~yr$^{-1}$ 
\citep[see][]{steven1}.\\

\noindent
{\bf \hi\ source catalogs.}\\
This data release includes 120 detections and 88 non-detections, for
which we provide upper limits below. The measured \hi\ parameters for the detected
galaxies are listed in Table~\ref{t_det}, ordered by increasing right ascension:\\

\bct
Cols. 1 and 2: GASS and SDSS identifiers. Galaxies with six digit GASS IDs are
part of GASS-low.\\

\bct
Col. 3: SDSS redshift, $z_{\rm SDSS}$. \\

\bct
Col. 4: on-source integration time of the Arecibo
observation, $T_{\rm on}$, in minutes. This number refers to
{\it on scans} that were actually combined, and does not account for
possible losses due to RFI excision (usually negligible). \\

\bct
Col. 5: velocity resolution of the final, smoothed spectrum in \kms. 
{\bc In general, lower signal-to-noise detections require more smoothing in order 
to better identify the edges and peaks of the \hi\ profiles, needed to measure
the \hi\ parameters.}\\

\bct
Col. 6: redshift, $z$, measured from the \hi\ spectrum.
The error on the corresponding heliocentric velocity, $cz$, 
is half the error on the width, tabulated in the following column.\\

\bct
Col. 7: observed velocity width of the source line profile
in \kms, \whi, measured at the 50\% level of each peak. 
{\bc Briefly, we fit straight lines to the sides of the \hi\ profile,
and identify the velocities $cz_r$, $cz_a$ corresponding to the
50\% peak flux density (from the fits) on the receding and approaching sides,
respectively (see Section 2.2 of \citealt{widths} for more details). The observed 
width is just the difference between these two velocities (and the \hi\ redshift
is given by their average).}
The error on the width is the sum in quadrature of the 
statistical and systematic uncertainties in \kms. Statistical errors
depend primarily on the signal-to-noise of the \hi\ spectrum, and are
obtained from the rms noise of the linear fits to the edges of the
\hi\ profile. Systematic errors depend on the subjective choice of the
\hi\ signal boundaries \citep[see][]{gass1}, and are negligible for most of
the galaxies in our sample (see also Appendix~B).\\

\bct
Col. 8: velocity width corrected for instrumental broadening
and cosmological redshift only, $W_{50}^c$, in \kms\ (see
\citealt{gass_dr2} for details). No inclination or turbulent motion
corrections are applied.\\

\bct
Col. 9: integrated \hi-line flux density in Jy \kms, $F_{\rm HI} \equiv \int S~dv$, measured on the 
smoothed and baseline-subtracted spectrum (observed velocity frame). The reported uncertainty 
is the sum in quadrature of the statistical and systematic errors (see col. 7).
The statistical errors are calculated according to equation 2 of S05
(which includes the contribution from uncertainties in the baseline fit).\\

\bct
Col. 10: rms noise of the observation in mJy, measured on the
signal- and RFI-free portion of the smoothed spectrum.\\

\bct
Col. 11: signal-to-noise ratio of the \hi\ spectrum, S/N,
estimated following \citet{saintonge07} and adapted to the velocity
resolution of the spectrum. 
This is the definition of S/N adopted by ALFALFA, which accounts for the
fact that for the same peak flux a broader spectrum has more signal.\\

\bct
Col. 12: base-10 logarithm of the \hi\ mass, \Mhi, in solar
units, computed via: 
\begin{equation}
    \frac{M_{\rm HI}}{\rm M_{\odot}} = \frac{2.356\times 10^5}{(1+z)^2}
    \left[ \frac{d_{\rm L}(z)}{\rm Mpc}\right]^2
    \left(\frac{\int S~dv}{\rm Jy~km~s^{-1}} \right)
\label{eq_MHI}
\end{equation}
\noindent
where $d_{\rm L}(z)$ is the luminosity distance to the galaxy at
redshift $z$ as measured from the \hi\ spectrum in the observed velocity frame \citep{danail_himass,meyer_himass}. \\

\bct
Col. 13: base-10 logarithm of the \hi\ mass fraction, \Mhi/\Mst.\\

\bct
Col. 14: quality flag, Q (1=good, 2=marginal and 5=confused). 
An asterisk indicates the presence of a note for the source in Appendix~B.
Code 1 indicates reliable detections, with a S/N ratio of order of
6.5 or higher. Marginal detections have lower S/N {\bc (between 5 and 6.5)}, thus more uncertain
\hi\ parameters, but are still secure detections, with \hi\ redshift
consistent with the SDSS one. We flag galaxies as ``confused'' when
most of the \hi\ emission is believed to originate from another source
within the Arecibo beam. For some of the galaxies, the presence of
small companions within the beam might contaminate (but is unlikely to
dominate) the \hi\ signal -- this is just noted in Appendix~B.\\

Table~\ref{t_ndet} gives the derived \hi\ upper limits for the non-detections. 
Columns 1-4 and 5 are the same as columns 1-4 and 10 in Table~\ref{t_det},
respectively. Column 6 lists the upper limit on the \hi\ mass in
solar units, $\log M_{{\rm HI},lim}$, computed assuming a 5$\sigma$ signal with 200 \kms\ 
velocity width, if the spectrum was smoothed to 100 \kms. Column 7
gives the corresponding upper limit on the gas fraction, log~\Mhi$_{,lim}$/\Mst.   
An asterisk in Column 8 indicates the presence of a note for the
galaxy in Appendix~B.\\

\begin{figure*}
\begin{center}
\includegraphics[width=15cm]{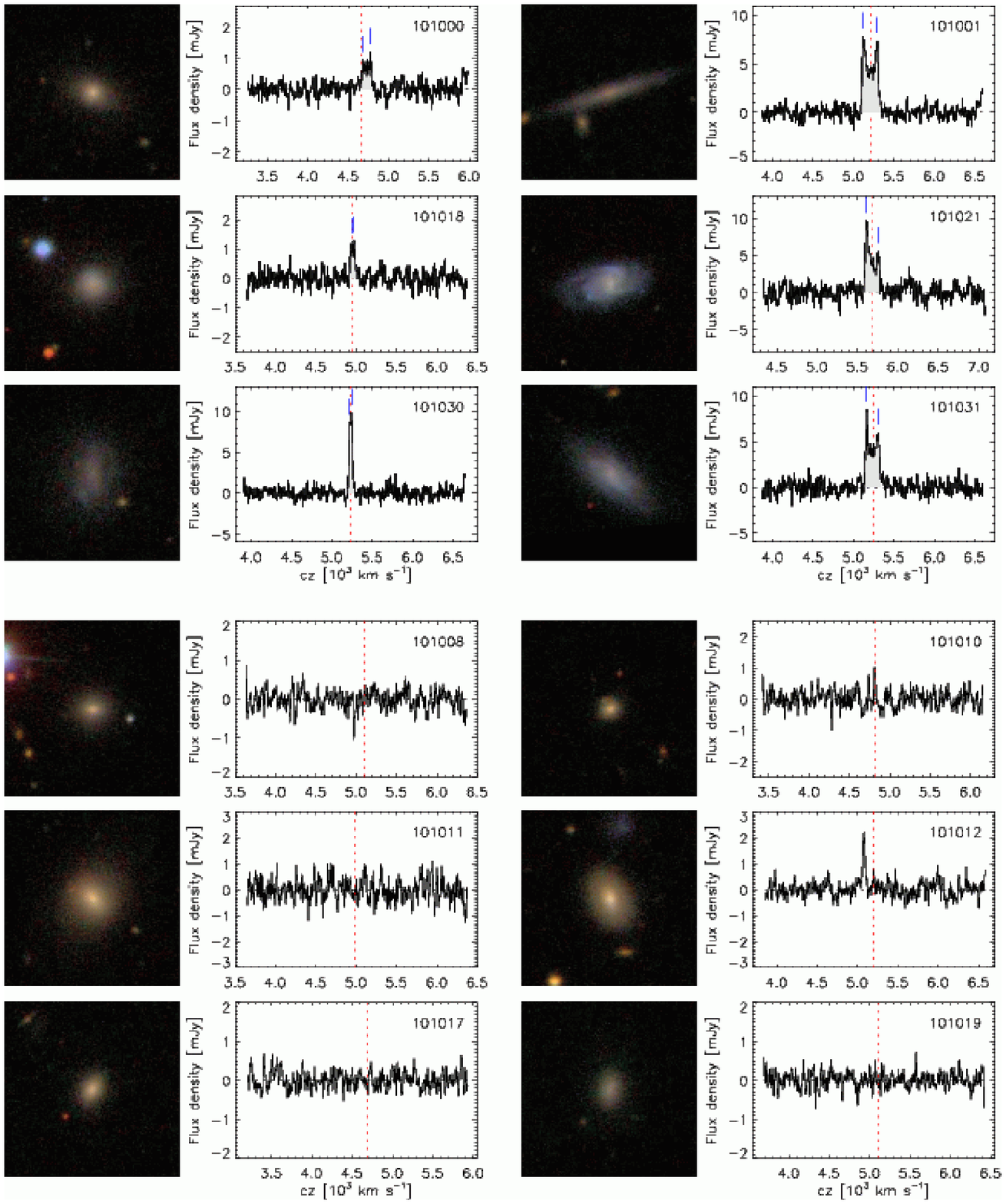}
\caption{SDSS postage stamp images (1 arcmin square) and
\hi-line profiles of GASS-low galaxies, ordered by increasing GASS number (indicated in each spectrum). 
{\it Top three rows:} \hi\ detections. The \hi\ spectra are
calibrated, smoothed and baseline-subtracted. A dotted line and two
dashes indicate the heliocentric velocity corresponding to the SDSS
redshift and the two peaks used for width measurement,
respectively. {\it Bottom three rows:} non-detections.
This is a sample of the complete figure, which is
available in the online version of the article.}
\label{spectra}
\end{center}
\end{figure*}

\noindent
{\bf SDSS postage stamps and \hi\ spectra.}\\
Figure~\ref{spectra} shows SDSS images and Arecibo \hi\ spectra for a subset
of galaxies included in this data release (top three rows: \hi\ detections;
bottom three rows: non-detections). 
The objects in each figure (detections and non-detections) are ordered by 
increasing GASS number, indicated on the top right corner of each spectrum.
The SDSS images show a 1 arcmin square field, \ie\ only the central
part of the region sampled by the Arecibo beam (the half
power full width of the beam is \about 3.5\arcmin\ at the
frequencies of our observations). Therefore, companions that might be
detected in our spectra typically are not visible in the
postage stamps, but they are noted in Appendix~B.
The \hi\ spectra are always displayed over a 3000 \kms\ velocity
interval, which includes the full 12.5 MHz bandwidth adopted for our
observations. The \hi-line profiles are calibrated, smoothed 
(to a velocity resolution between 5 and 15 \kms\ for
the detections, as listed in Table~\ref{t_det}, or to 15 \kms\ for the non-detections), and
baseline-subtracted. A red, dotted line indicates the heliocentric
velocity corresponding to the optical redshift from SDSS. 
For the \hi\ detections, the shaded area and two vertical
dashes show the part of the profile that was integrated to
measure the \hi\ flux and the peaks used for width measurement,
respectively.

\section*{Appendix B: Notes on Individual Objects}\label{s_notes}

We list here notes for galaxies marked with an asterisk in
the last column of Tables~\ref{t_det} and \ref{t_ndet}.
The galaxies are ordered by increasing GASS number. In what follows, 
AA2 is the abbreviation for ALFALFA detection code 2.\\


\noindent
{\bf Detections (Table \ref{t_det})}\\

\noindent
{\bf 101016}  --  asymmetric profile, uncertain width; confused with NGC 675 ($cz=5335$ \kms\ from NED) \about 2.5 arcmin W; also notice 
                  large elliptical \about 1 arcmin W (NGC 677, 5082 \kms). Several small blue galaxies around, $z>0.08$. \\
{\bf 101018}  --  {\bc polarization mismatch}. \\
{\bf 101021}  --  RFI spike at 1392 MHz (\about 6120 \kms); early-type galaxy \about 1 arcmin SE has $z=0.048$. \\
{\bf 101030}  --  several galaxies within 3 arcmin, $z>0.06$. \\
{\bf 107019}  --  small blue galaxy \about 1 arcmin W, $z=0.105$. \\
{\bf 108011}  --  blue galaxy \about 1.5 arcmin SW, $z=0.052$. \\
{\bf 108014}  --  blend with large, edge-on disk \about 2 arcmin SE (SDSS J081011.20+245334.7, $z=0.014056$); the face-on spiral 
                  \about 3 arcmin S haz $z=0.081$. \\
{\bf 108019}  --  blend with blue companion \about 2 arcmin W (SDSS J081002.90+224623.1, $z=0.015736$, $cz=4718$ \kms). \\
{\bf 108024}  --  confused/blend with two large spirals \about 1 arcmin NW (SDSS J082401.55+210138.3, $z=0.015483$) 
                  and \about 2.5 arcmin SW (SDSS J082355.28+205831.6, $z=0.015708$). \\
{\bf 108029}  --  blend; interacting with large, blue companion 1.8 arcmin N (UGC 4264, $z=0.013642$). \\
{\bf 108049}  --  AA2. \\
{\bf 108051}  --  AA2. \\
{\bf 108078}  --  small blue galaxy \about 1 arcmin W has no redshift; blue galaxy \about 1.5 arcmin S has $z=0.07$. \\
{\bf 108093}  --  blue spiral \about 2 arcmin W, SDSS J083520.21+233943.2, has $z=0.039$. \\
{\bf 108097}  --  small blue galaxy \about 3 arcmin W, $z=0.130$. \\
{\bf 108129}  --  two small blue galaxies \about 2 arcmin W, $z>0.06$. \\
{\bf 108140}  --  a few small galaxies within 3.5 arcmin, all in the background or without optical redshift; AA2. \\
{\bf 108143}  --  face-on companion \about 2.5 arcmin N (SDSS J080037.51+134150.4, $z=0.01525$, $cz=4572$ \kms) not detected; blue galaxy \about 40 arcsec W,
                  SDSS J080035.43+133936.8, has $z=0.073$. \\
{\bf 108145}  --  {\bc polarization mismatch}; high-frequency edge uncertain, systematic error. Companion \about 2 arcmin W 
                  (SDSS J080158.07+092324.2, $z=0.013337$, $cz=3998$ \kms) and two early-types \about 1.5 arcmin SE (SDSS J080210.94+092141.7, $cz=4795$ \kms)
                  and \about 2 arcmin E (NGC 2511, $cz=4467$ \kms\ from NED) not detected. \\
{\bf 109009}  --  near bright star; AA2.  \\
{\bf 109020}  --  small blue galaxy \about 2 arcmin NE, $z=0.13$; AA2. \\
{\bf 109034}  --  high-frequency edge uncertain, systematic error. \\
{\bf 109058}  --  galaxy \about 2.5 arcmin SE has $z=0.132$. \\
{\bf 109077}  --  AA2. \\
{\bf 109079}  --  large, face-on blue companion \about 1.5 arcmin NE (UGC 5344, $cz$\about 4135 \kms) also detected. \\
{\bf 109083}  --  AA2. \\
{\bf 109094}  --  small blue galaxy \about 2.5 arcmin W, $z=0.012$. \\
{\bf 109108}  --  AA2. \\
{\bf 109120}  --  small galaxy \about 1.5 arcmin W, $z=0.09$; AA2. \\
{\bf 109126}  --  AA2. \\
{\bf 109129}  --  blend with blue companion \about 1 arcmin E (SDSS J090133.86+123931.6, $z=0.019761$). \\
{\bf 109135}  --  AA2. \\
{\bf 110013}  --  AA2. \\
{\bf 110019}  --  galaxy \about 1.5 arcmin SW has $z=0.143$. \\
{\bf 110038}  --  264 mJy continuum source at 2 arcmin, but very strong signal. \\
{\bf 110054}  --  blue, edge-on companion \about 3.5 arcmin SE (SDSS J104410.01+221233.2, $z=0.019602$), unlikely to contribute significantly 
                  to the signal; early-type galaxy \about 1 arcmin W has $z=0.110$. \\
{\bf 110057}  --  blend with large, blue companion \about 2 arcmin NW (SDSS J102916.83+260557.2, $z=0.01699$). \\
{\bf 110080}  --  two blue disks \about 2.5 arcmin NW and \about 2.5 arcmin N have redshifts $z=0.046$ and $z=0.162$, respectively. \\
{\bf 111004}  --  blend with spectacular blue companion \about 3 arcmin E (SDSS J115805.22+275243.8, $z=0.011149$). \\
{\bf 111047}  --  small blue galaxy \about 0.5 arcmin NE, no optical $z$. \\
{\bf 110058}  --  blend with blue companion \about 3 arcmin N (SDSS J101823.53+131642.1, $z=0.018273$). \\
{\bf 111053}  --  edge-on galaxy \about 2 arcmin NW, $z=0.067$. \\
{\bf 111063}  --  {\bc polarization mismatch}; large blue galaxy \about 4 arcmin NW (NGC 4005, $cz=4469$ \kms) not detected; blue disk 
                  \about 4 arcmin W (SDSS J115809.49+250520.0, $z=0.014286$, 4283 \kms), significant contamination unlikely. \\
{\bf 111086}  --  blend with blue companion \about 3 arcmin N (SDSS J113913.53+150215.7, $z=0.014011$). \\
{\bf 112035}  --  early-type companion \about 2 arcmin SE, SDSS J125910.30+273711.5 ($z=0.01916$, 5744 \kms); AA2. \\
{\bf 112068}  --  high-frequency edge uncertain, systematic error. Large, red disk \about 1.5 arcmin NW (NGC 4063, $cz=4917$ \kms) not
                  detected; disk \about 2.5 arcmin E has $z=0.024$.  \\
{\bf 112080}  --  blend: most of the \hi\ signal comes from large, blue companion \about 2 arcmin E (NGC 4615, $z=0.015731$ from NED); 
                  also notice spiral galaxy \about 3 arcmin S at the same redshift (SDSS J124131.46+260233.5, $z=0.015957$). \\
{\bf 112014}  --  confused with face-on, blue companion \about 20 arcsec W (SDSS J120429.88+022654.5, $z=0.020141$, 6038 \kms); blue 
                  galaxy \about 2 arcmin W (SDSS J120424.60 +022732.8, $cz=6730$ \kms) not detected. \\
{\bf 112112}  --  two small galaxies \about 1.5 and \about 2 arcmin NE, no $z$. \\
{\bf 112116}  --  galaxies \about 1 arcmin SW and 2 arcmin S in the background ($z=0.021$ and $z=0.025$ respectively). \\
{\bf 113010}  --  RFI spike at 1408 MHz (\about 2650 \kms).  \\
{\bf 113011}  --  disk galaxy \about 2.5 arcmin NE has $z=0.112$. \\
{\bf 113038}  --  tiny blue comp. \about 2.5 arcmin SW, SDSS J132408.93+355317.6 ($z=0.018561$). \\
{\bf 113091}  --  3 small blue galaxies within 3 arcmin, $z>0.04$. \\
{\bf 113115}  --  small blue galaxies \about 1.5 arcmin N and \about 3 arcmin SW in the background ($z>0.04$). \\
{\bf 113118}  --  two galaxies \about 2.5 arcmin SW and \about 3 arcmin NW have $z=0.06$. \\
{\bf 113122}  --  low-frequency edge uncertain, systematic error. Several galaxies within 3 arcmin, $z>0.05$. \\
{\bf 113123}  --  blue disk galaxy \about 4 arcmin NE, $z=0.024$. \\
{\bf 113124}  --  small blue galaxy \about 2 arcmin NE, $z=0.06$. Unclear what is causing the feature at 1398.5-99 MHz (\about 4600 \kms), 
                  but it is present in both polarizations. \\
{\bf 114001}  --  small blue galaxy \about 45 arcsec W, no optical $z$. \\
{\bf 114005}  --  confused/blend with blue companion \about 2 arcmin NW (IC 1013, $z=0.015341$, 4599 \kms);
                  3 large galaxies within 3 arcmin at the same redshift ($z=0.015$), and a smaller one 2 arcmin E ($z=0.013$). \\
{\bf 114033}  --  blend with two large, blue companions \about 1 arcmin NE (SDSS J141216.00+155247.4, $z=0.017549$) and \about 2 arcmin SE
                  (NGC 5504, $z=0.017505$ from NED). \\
{\bf 114044}  --  profile edges uncertain, systematic error. Small blue galaxy \about 45 arcsec SW, no optical $z$. \\
{\bf 114047}  --  blend/confused with blue triplet \about 2.5 arcmin E (SDSS J140520.30+102438.6, $z=0.018738$; 
                  SDSS J140519.91+102445.6, no $z$; SDSS J140520.30+102452.9, $z=0.018332$); also notice large
                  early-type \about 2 arcmin E (SDSS J140516.52+102551.7, $z=0.017949$). \\
{\bf 114076}  --  small blue galaxy \about 1.5 arcmin N, no $z$; AA2. \\
{\bf 114077}  --  early-type galaxy \about 2 arcmin E, $z=0.030$; edge-on disk \about 0.5 arcmin W, no optical $z$. \\
{\bf 114091}  --  large spiral \about 1.5 arcmin NE (UGC 9165, $cz=5259$ \kms\ from NED) also detected;
                  face-on spiral \about 2.5 arcmin SW has $z=0.079$. Absorption-like feature at \about 5050 \kms\ is not RFI and 
                  is present in both polarizations (nothing obvious in the {\it off} position, just a few small blue galaxies without redshifts). \\
{\bf 114110}  --  blend/confused with large blue companion \about 20 arcsec S, NGC 5491 ($z=0.019647$ from NED). \\
{\bf 114115}  --  small blue galaxy \about 3.5 arcmin NE, $z=0.025$. \\
{\bf 114144}  --  120 mJy continuum source at 2 arcmin, standing waves. AA2. \\
{\bf 122001}  --  two large galaxies \about 2 arcmin NE and 3 arcmin E have $z=0.037$; AA2. \\
{\bf 122002}  --  strong signal detected on top of long standing wave (in polarization A). \\
{\bf 124002}  --  AA2. \\
{\bf 124012}  --  galaxy \about 1.5 arcmin NE, $z=0.15$; AA2. \\


\noindent
{\bf Non-detections (Table \ref{t_ndet})}\\

\noindent
{\bf 101008}  --  blue disk \about 2 arcmin N has $z=0.034$. \\
{\bf 101012}  --  detected companion ($cz=5080$ \kms), most likely the blue smudge \about 10 arcsec N (SDSS J014902.23+125603.2, no optical redshift).  \\
{\bf 108013}  --  galaxy \about 3 arcmin SW, SDSS J080229.78 +092743.2 ($z=0.014515$, 4351 \kms), also not detected. \\
{\bf 108022}  --  blue companion \about 1 arcmin NW (SDSS J080206.54+092238.3, $z=0.013795$), also not detected. \\
{\bf 108080}  --  blue disk galaxy \about 2 arcmin W has $z=0.047$. \\
{\bf 108101}  --  perhaps {\bc hint of \hi\ galaxy signal.} \\
{\bf 108114}  --  blue companion \about 3.5 arcmin N (SDSS J081919.78+210331.8, $z=0.016114$), also not detected. \\
{\bf 109022}  --  several galaxies within 3 arcmin, all in the background or without optical redshift. \\
{\bf 109036}  --  AA2. \\
{\bf 109064}  --  detected large spiral \about 2 arcmin S, (NGC 2874, $z=0.012573$, 3769 \kms); also notice large elliptical 
                  \about 2 arcmin SW, (NGC 2872, 3196 \kms\ from NED). \\
{\bf 109075}  --  low surface brightness galaxy \about 40 arcsec W, no optical $z$; 4 other galaxies within 3-4 arcmin, $z>0.06$. \\
{\bf 109132}  --  detected blue, face-on companion \about 1.5 arcmin SE, SDSS J095905.12+102140.0 ($z=0.017938$, 5378 \kms). \\
{\bf 110042}  --  several galaxies within 3 arcmin, all in the background or without optical redshift. \\
{\bf 110047}  --  two galaxies within \about 1 arcmin E without optical $z$, probably in background. \\
{\bf 110060}  --  large, early-type companion \about 3 arcmin S, SDSS J102058.56+253109.8 ($z=0.020785$, 6231 \kms), also not detected. \\
{\bf 111024}  --  large, early-type companion \about 1.5 arcmin E, SDSS J114428.32+163329.1 ($z=0.011058$, 3315 \kms), also not detected. \\
{\bf 111049}  --  detected blue companions \about 1.5 arcmin SW, SDSS J115752.02+250254.1 ($z=0.014196$, 4256 \kms)
                  and \about 3 arcmin E, SDSS J115809.49+250520.0 ($z=0.014286$, 4283 \kms); also notice large spiral
                  \about 4 arcmin NE, NGC 4005 (4464 \kms). \\
{\bf 111066}  --  {\bc hint of \hi\ galaxy signal}; blue face-on disk \about 2.5 arcmin SW has $z=0.085$. \\
{\bf 112002}  --  galaxy \about 3 arcmin NE, SDSS J125806.06+272508.1 ($z=0.019237$, 5767 \kms) also not detected; galaxies \about 2.5 arcmin NE and 
                  \about 4 arcmin NW are in the background.  \\
{\bf 112003}  --  spiral galaxy \about 2 arcmin E has $z=0.025$; small galaxies \about 3 arcmin S also in background. \\
{\bf 112004}  --  large disk \about 1 arcmin SW, SDSS J125746.16+274525.3 ($z=0.02051$, 6149 \kms) and small galaxy \about 2 arcmin SE, SDSS J125752.31+274422.7 
                  ($z=0.023039$, 6907 \kms) also not detected; 4 other galaxies within 3 arcmin, $z>0.07$. \\
{\bf 112011}  --  several galaxies within 3 arcmin, all in the background or without optical $z$. \\
{\bf 112012}  --  galaxy \about 2.5 arcmin N has $z=0.023$. \\
{\bf 112016}  --  perhaps {\bc hint of \hi\ galaxy signal} (offset from SDSS redshift); small, blue companion \about 1.5 arcmin SW (SDSS J125317.61+262021.4, 
                  $z=0.02311$, 6928 \kms). \\
{\bf 112017}  --  crowded field. \\
{\bf 112060}  --  two nearby galaxies in background (\about 2 arcmin W, $z=0.025$ and \about 3 arcmin N, $z=0.026$). \\
{\bf 112065}  --  perhaps {\bc hint of \hi\ signal} (from galaxy and/or blue companion \about 3.5 arcmin SW, SDSS J120055.36+152221.8, $z=0.017482$, 5241 \kms). \\
{\bf 112084}  --  detected face-on, blue companion \about 1.5 arcmin NW, SDSS J125905.29+273839.9 ($z=0.018117$,
                  5431 \kms); also notice smaller blue companion \about 2 arcmin SE (SDSS J125918.54+273536.9, $z=0.017777$, 5329 \kms) and 
                  galaxy \about 3 arcmin W (SDSS J125858.10+273540.9, $z=0.020057$, 6013 \kms), both not detected. \\
{\bf 112089}  --  detected blue companion \about 1.5 arcmin NW, SDSS J125020.21+264459.4 ($z=0.02373$, 7114 \kms); perhaps {\bc hint of \hi\ signal} from
                  disk galaxy \about 1.5 arcmin S, SDSS J125026.59+264232.3 ($z=0.018702$, 5607 \kms). \\
{\bf 112106}  --  detected large blue companion \about 1 arcmin NE (UGC 7035, $z=0.00409$) in board 4, \about 1414.5 MHz. \\
{\bf 113001}  --  galaxy \about 1.5 arcmin NE, SDSS J130416.46+273022.9, has slightly higher redshift ($z=0.024$, \about 7200 \kms). \\
{\bf 113004}  --  large face-on spiral \about 3 arcmin SE (SDSS J130126.12+275309.5, $z=0.018271$, 5478 \kms) and small galaxy \about 2 arcmin SW 
                  (SDSS J130104.82+275330.3, $z=0.019293$, 5784 \kms) also not detected. \\
{\bf 113012}  --  over ten galaxies within 3 arcmin at slightly higher redshift ($z>0.018$). \\
{\bf 113025}  --  four small galaxies within 1 arcmin, all without optical redshifts and most likely in the background. \\
{\bf 113032}  --  early-type companion \about 3 arcmin SE, SDSS J130019.10+273313.3 ($z=0.019644$, 5889 \kms) also not detected;
                  two galaxies \about 3 arcmin E and \about 2 arcmin W in background. \\
{\bf 113040}  --  two galaxies within \about 1 arcmin E, $z>0.025$; blue smudge \about 20 arcsec S, no $z$. \\
{\bf 113047}  --  companion \about 3 arcmin NW, SDSS J130010.41+273542.0 ($z=0.018676$, 5600 \kms);
                  three galaxies between 1 and 3 arcmin SE, $z=0.024-0.026$. \\
{\bf 113051}  --  four galaxies within \about 3 arcmin, $z=0.022-0.028$. \\
{\bf 113060}  --  crowded field. \\
{\bf 113078}  --  early-type \about 1 arcmin SW, SDSS J130231.87\- +275607.9 ($z=0.022091$, 6623 \kms) also not detected. \\
{\bf 113100}  --  {\bc hint of \hi\ galaxy signal}; AA2. Three small galaxies 2-4 arcmin away, $z=0.02-0.025$. \\
{\bf 113128}  --  crowded field. \\
{\bf 114008}  --  marginally detected companion \about 1.7 arcmin W, GASS 114005 ($z=0.015153$, 4543 \kms; blend, see notes above); also 
                  notice SDSS J142807.22+255207.5 ($z=0.014591$, 4374 \kms) and NGC 5629 ($cz=4498$ \kms\ from NED). \\
{\bf 114025}  --  detected blue companion \about 2.5 arcmin E, SDSS J142532.21+254300.1 ($z=0.013609$, 4080 \kms); the disk
                  galaxy \about 1 arcmin S of the blue companion has $z=0.075$. \\
{\bf 114036}  --  blue companion 2.5 arcmin NE, SDSS J142758.85+255158.7 ($z=0.015341$, 4599 \kms), also not detected;
                  notice two early-type galaxies \about 2 arcmin N ($z=0.014$) and \about 3 arcmin E ($z=0.015$). \\
{\bf 114037}  --  detected companion, probably small blue galaxy \about 1.5 arcmin NW (SDSS J142829.10+272113.4, no $z$); 
                  NGC 5635 \about 4 arcmin N ($cz=4316$ \kms\ from NED). \\
{\bf 114038}  --  small galaxy 2 arcmin E, SDSS J142723.43 +255240.2 ($z=0.013418$, 4023 \kms). \\
{\bf 114065}  --  early-type companion \about 2 arcmin W, SDSS J142714.65+255319.1 ($z=0.0159$, 4767 \kms). \\
{\bf 114096}  --  a few galaxies between 2 and 3 arcmin in background, in particular SDSS J142544.50+254636.0, a blue 
                  galaxy 2 arcmin W ($z=0.075$) and SDSS J142600.40+254320.3, a large early-type \about 3 arcmin S ($z=0.076$ from NED). \\
{\bf 123011}  --  three galaxies between 2 and 3 arcmin, $z>0.04$. \\

\onecolumn
\begin{landscape}
\begin{table*}
\small
\centering
\caption{SDSS and UV parameters of GASS-low galaxies}
\label{t_sdss}
\begin{tabular}{rclcccccccccccc}
\hline\hline
   &  &  &  & log \Mst & \Rinz & $R_{50}$ & $R_{90}$ & log \must  & ext$_r$ & r & (b/a)$_r$ & incl &\nuvr & SFR  \\
GASS  & SDSS ID & Other name & $z_{\rm SDSS}$ & (\Msun) & (\arcsec) & (\arcsec)&(\arcsec) & (\Msun~kpc$^{-2}$)& (mag)& (mag) &   & (deg) & (mag)& (\Msun~yr$^{-1}$) \\
(1)  & (2)  & (3)  & (4)  & (5) & (6) & (7) & (8)  & (9)  &  (10) & (11) & (12) & (13) & (14) & (15) \\
\hline
124009 & J000619.61+141938.7 &  --          & 0.0182 &  9.66 &  3.52 &  3.82 & 11.17 &  8.63 &  0.31 & 15.47 &   0.394 &   70 &  5.27 &   0.078 \\
124012 & J000629.29+141056.5 &  --          & 0.0178 &  9.74 &  6.03 &  6.72 & 16.28 &  8.27 &  0.34 & 15.79 &   0.144 &   90 &  4.26 &   0.162 \\
124006 & J001947.33+003526.8 &  --          & 0.0177 &  9.75 &  3.47 &  3.36 & 10.58 &  8.76 &  0.07 & 14.75 &   0.753 &   42 &  3.82 &   0.539 \\
124004 & J002534.40+005048.6 &  --          & 0.0178 &  9.31 &  5.35 &  6.11 & 11.54 &  7.93 &  0.06 & 15.32 &   0.917 &   24 &  2.21 &   0.326 \\
124002 & J004903.69+152907.9 &  --          & 0.0183 &  9.25 &  2.60 &  2.57 &  6.02 &  8.48 &  0.17 & 15.72 &   0.479 &   64 &  2.33 &   0.209 \\
101021 & J011653.58+000911.2 &  --          & 0.0190 &  9.22 &  4.22 &  5.03 & 11.03 &  8.00 &  0.09 & 15.49 &   0.686 &   48 &  2.21 &   0.276 \\
101031 & J014755.16+124131.0 &  --          & 0.0175 &  9.17 &  5.51 &  6.56 & 14.74 &  7.79 &  0.17 & 15.56 &   0.492 &   63 &  2.41 &   0.154 \\
101030 & J014803.60+125604.6 &  --          & 0.0175 &  9.10 &  7.06 &  8.85 & 17.38 &  7.50 &  0.18 & 15.99 &   0.659 &   50 &  2.90 &   0.061 \\
101000 & J014853.12+132526.2 &  --          & 0.0155 &  9.41 &  4.28 &  3.99 & 12.06 &  8.35 &  0.18 & 15.86 &   0.787 &   39 &  4.91 &   0.096 \\
101012 & J014902.52+125539.0 &  --          & 0.0174 &  9.83 &  3.77 &  4.01 &  9.92 &  8.78 &  0.22 & 15.06 &   0.642 &   52 &  5.17 &   0.027 \\
101024 & J014917.63+132759.9 &  --          & 0.0166 &  9.28 &  3.81 &  4.18 &  9.82 &  8.26 &  0.21 & 16.17 &   0.735 &   44 &  5.57 &   0.067 \\
101016 & J014918.93+130252.0 &  --          & 0.0178 &  9.72 &  2.53 &  2.74 &  6.50 &  8.99 &  0.24 & 15.53 &   0.845 &   33 &  5.51 &   0.007 \\
101019 & J014920.31+131754.5 &  --          & 0.0171 &  9.06 &  4.78 &  4.57 & 11.24 &  7.83 &  0.21 & 16.68 &   0.673 &   49 &  3.92 &   0.092 \\
101011 & J014933.52+131400.0 &  --          & 0.0167 &  9.95 &  4.66 &  4.96 & 14.21 &  8.75 &  0.20 & 14.84 &   0.768 &   41 &  5.92 &   0.013 \\
101017 & J014953.83+125833.7 &  --          & 0.0156 &  9.35 &  2.92 &  2.94 &  8.24 &  8.62 &  0.24 & 16.01 &   0.657 &   50 &  5.22 &   0.009 \\

\hline
\end{tabular}
\begin{flushleft}
Note. -- The full version of this table is available online.
\end{flushleft}
\end{table*}
\end{landscape}

\begin{landscape}
\begin{table*}
\small
\centering
\caption{\hi\ Properties of GASS-low detections}
\label{t_det}
\begin{tabular}{cccccccccccccl}
\hline\hline
      &         &               & $T_{\rm on}$ & $\Delta v$ &     & \whi  & $W_{50}^c$&  \Fhi      &  rms & &log \Mhi  &   & \\
GASS  & SDSS ID & $z_{\rm SDSS}$ & (min)       &  (\kms)    & $z$ &  (\kms)& (\kms) & (Jy \kms) & (mJy)& S/N  & (\Msun) & log \Mhi/\Mst & Q \\
(1)  & (2)  & (3)  & (4)  & (5)  & (6)  & (7)  & (8)  & (9)  &  (10) & (11) & (12) & (13) & (14)\\
\hline
124012 & J000629.29+141056.5 & 0.0178 &   5 &  15 & 0.017\,712 & 267$\pm$   5 & 256 &  0.65$\pm$  0.10 &  0.64 &  11.4 &  8.94 & $-$0.81 &  1*  \\
124004 & J002534.40+005048.6 & 0.0178 &   5 &  12 & 0.017\,816 & 130$\pm$   2 & 122 &  0.71$\pm$  0.07 &  0.72 &  17.5 &  8.99 & $-$0.32 &  1	\\
124002 & J004903.69+152907.9 & 0.0183 &   4 &  12 & 0.018\,269 & 205$\pm$   3 & 195 &  0.67$\pm$  0.11 &  0.90 &  10.6 &  8.98 & $-$0.26 &  1*  \\
101021 & J011653.58+000911.2 & 0.0190 &   5 &  12 & 0.018\,956 & 180$\pm$   4 & 170 &  0.93$\pm$  0.10 &  0.88 &  15.9 &  9.15 & $-$0.06 &  1*  \\
101031 & J014755.16+124131.0 & 0.0175 &   5 &  12 & 0.017\,465 & 173$\pm$   3 & 164 &  0.86$\pm$  0.07 &  0.67 &  19.7 &  9.05 & $-$0.12 &  1	\\
101030 & J014803.60+125604.6 & 0.0175 &  10 &  10 & 0.017\,445 &  44$\pm$   1 &  39 &  0.42$\pm$  0.03 &  0.57 &  25.0 &  8.74 & $-$0.36 &  1*  \\
101000 & J014853.12+132526.2 & 0.0155 &  59 &  12 & 0.015\,768 & 115$\pm$   3 & 107 &  0.09$\pm$  0.02 &  0.21 &   8.6 &  8.01 & $-$1.40 &  1	\\
101016 & J014918.93+130252.0 & 0.0178 &  50 &  15 & 0.017\,242 & 207$\pm$  48 & 196 &  0.40$\pm$  0.03 &  0.20 &  24.5 &  8.70 & $-$1.01 &  5*  \\
101001 & J015032.27+133942.2 & 0.0174 &  10 &  12 & 0.017\,372 & 202$\pm$   5 & 193 &  1.10$\pm$  0.07 &  0.56 &  27.9 &  9.15 & \phm 0.07 &  1	\\
101018 & J015036.88+130636.8 & 0.0165 &  59 &  12 & 0.016\,531 &  69$\pm$   7 &  62 &  0.07$\pm$  0.02 &  0.24 &   7.5 &  7.94 & $-$1.31 &  1*  \\
107019 & J074842.59+263223.2 & 0.0155 &  25 &  12 & 0.015\,537 & 172$\pm$   3 & 164 &  0.22$\pm$  0.04 &  0.33 &  10.1 &  8.35 & $-$0.81 &  1*  \\
107026 & J075433.01+294238.8 & 0.0167 &   5 &  10 & 0.016\,672 & 223$\pm$   3 & 215 &  3.58$\pm$  0.11 &  1.02 &  53.3 &  9.63 & \phm 0.26 &  1	\\
108143 & J080038.05+133923.0 & 0.0158 &  23 &  12 & 0.015\,754 & 101$\pm$   7 &  94 &  0.26$\pm$  0.04 &  0.51 &  10.3 &  8.45 & $-$0.68 &  1*  \\
108136 & J080048.38+100500.9 & 0.0137 &  35 &  12 & 0.013\,696 & 212$\pm$  11 & 203 &  0.28$\pm$  0.04 &  0.37 &  10.5 &  8.35 & $-$1.12 &  1	\\
108145 & J080206.54+092238.3 & 0.0138 &  20 &  15 & 0.013\,736 &  91$\pm$  10 &  82 &  0.10$\pm$  0.03 &  0.30 &   6.3 &  7.90 & $-$1.26 &  2*  \\
\hline	
\end{tabular}
\begin{flushleft}
Note. -- The full version of this table is available online.
\end{flushleft}
\end{table*}
\end{landscape}

\begin{table*}
\small
\centering
\caption{GASS-low non-detections}
\label{t_ndet}
\begin{tabular}{ccccccccc}
\hline\hline
      &         &              &  $T_{\rm on}$ &  rms  & log \Mhi$_{,lim}$ &    &   \\
GASS  & SDSS ID & $z_{\rm SDSS}$& (min)        &  (mJy)& (\Msun)  & log \Mhi$_{,lim}$/\Mst & Note  &   \\
(1)  & (2)  & (3)  & (4)  & (5)  & (6)  & (7) & (8)  &   \\
\hline
124009 & J000619.61+141938.7 & 0.0182 &  90 &  0.16 &  7.95 &  $-$1.70   &  ... \\
124006 & J001947.33+003526.8 & 0.0177 &  65 &  0.21 &  8.05 &  $-$1.71   &  ... \\
101012 & J014902.52+125539.0 & 0.0174 &  35 &  0.28 &  8.14 &  $-$1.68   &  *	\\
101024 & J014917.63+132759.9 & 0.0166 &  45 &  0.23 &  8.03 &  $-$1.25   &  ... \\
101019 & J014920.31+131754.5 & 0.0171 &  68 &  0.19 &  7.97 &  $-$1.09   &  ... \\
101011 & J014933.52+131400.0 & 0.0167 &  20 &  0.39 &  8.26 &  $-$1.69   &  ... \\
101017 & J014953.83+125833.7 & 0.0156 &  48 &  0.22 &  7.95 &  $-$1.40   &  ... \\
101008 & J014954.10+130735.9 & 0.0170 &  48 &  0.21 &  8.02 &  $-$1.33   &  *	\\
101010 & J015015.67+130826.8 & 0.0161 &  39 &  0.26 &  8.04 &  $-$1.13   &  ... \\
108111 & J080116.61+091553.3 & 0.0155 &  10 &  0.43 &  8.24 &  $-$1.72   &  ... \\
108080 & J080134.48+091817.4 & 0.0153 &  36 &  0.30 &  8.07 &  $-$1.25   &  *	\\
108065 & J080158.49+150331.5 & 0.0161 &   5 &  0.65 &  8.45 &  $-$1.65   &  ... \\
108022 & J080210.94+092141.8 & 0.0160 &   8 &  0.63 &  8.43 &  $-$1.61   &  *	\\
108052 & J080212.07+103234.1 & 0.0145 &  30 &  0.27 &  7.97 &  $-$1.63   &  ... \\
108013 & J080237.85+093000.3 & 0.0159 &  10 &  0.59 &  8.40 &  $-$1.69   &  *	\\
\hline
\end{tabular}
\begin{flushleft}
Note. -- The full version of this table is available online.
\end{flushleft}
\end{table*}

\twocolumn

\end{document}